\documentclass[fleqn,usenatbib]{mnras}

\usepackage{newtxtext,newtxmath}
\usepackage[T1]{fontenc}
\usepackage{ae,aecompl}

\usepackage{graphicx}	% Including figure files
\usepackage{amsmath}	% Advanced maths commands
\usepackage{amssymb}	% Extra maths symbols

\title[Resolving accretion flows in AGN]{Resolving accretion flows in nearby active galactic nuclei with the Event Horizon Telescope}

\author[B. Bandyopadhyay et al.]{
Bidisha Bandyopadhyay,$^{1}$\thanks{E-mail: bidisharia@gmail.com (BB)}
Fu-Guo Xie,$^{2}$
Neil M. Nagar,$^{1}$
Dominik R. G. Schleicher,$^{1}$
\newauthor 
Venkatessh Ramakrishnan,$^{1}$
Patricia Ar\'evalo,$^{3}$
Elena L\'opez,$^{3}$
and Yaherlyn Diaz$^{3}$
\\
$^{1}$Departamento de Astronom\'ia, Facultad Ciencias F\'isicas y Matem\'aticas, Universidad de Concepci\'on, \\
Av. Esteban Iturra s/n Barrio Universitario, Casilla 160-C, Concepci\'on, Chile\\
$^{2}$Key Laboratory for Research in Galaxies and Cosmology, Shanghai Astronomical Observatory, Chinese Academy of Sciences, \\
80 Nandan Road, Shanghai 200030, China\\
$^{3}$Instituto de F\'isica y Astronom\'ia, Facultad de Ciencias, Universidad de Valpara\'iso, Gran Bretana No. 1111, Playa Ancha, \\
Valpara\'iso, Chile
}

\date{Accepted XXX. Received YYY; in original form ZZZ}

\pubyear{2019}

\begin{document}
\label{firstpage}
\pagerange{\pageref{firstpage}--\pageref{lastpage}}
\maketitle

\begin{abstract}
The Event Horizon Telescope (EHT), now with its first ever image of the photon ring around the supermassive black hole of M87, provides a unique opportunity to probe the physics of supermassive black holes through Very Long Baseline Interferometry (VLBI), such as the existence of the event horizon, the accretion processes as well as jet formation in Low Luminosity AGN (LLAGN). We build a theoretical model which includes an Advection Dominated Accretion Flow (ADAF) with emission from thermal and non-thermal electrons in the flow and a simple radio jet outflow. The predicted spectral energy distribution (SED) of this model is compared to sub-arcsec resolution observations to get the best estimates of the model parameters. The model-predicted radial emission profiles at different frequency bands are used to predict whether the inflow can be resolved by the EHT or with telescopes such as the Global 3-mm VLBI array (GMVA). In this work the model is initially tested with high resolution SED data of M87 and then applied to our sample of 5 galaxies (Cen A, M84, NGC 4594, NGC 3998 and NGC 4278). The model then allows us to predict if one can detect and resolve the inflow for any of these galaxies using the EHT or GMVA within an 8 hour integration time.   
\end{abstract}


%%%%%%%%%%%%%%%%%%%%%%%%%%%%%%%%%%%%%%%%%%%%%%%%%%

%%%%%%%%%%%%%%%%% BODY OF PAPER %%%%%%%%%%%%%%%%%%

\section{Introduction}
With the first results of the Event Horizon Telescope (EHT)\footnote{https://eventhorizontelescope.org/} \citep{Akiyama2019a, Akiyama2019b, Akiyama2019c, Akiyama2019d} showing a detection of the photon ring around the black hole in the nucleus of M87 (the central galaxy of the Virgo cluster), a new window has opened to probe regions in the extreme proximity of supermassive black hole. Such an advancement in science and technology has not only left a mark in testing Einstein's theory of general relativity (GR), but has also enabled us to probe regions in the accretion flow which were previously unresolvable. It is important to investigate the different physical processes that occur in the accretion flow to gain an insight on the source of energy powering such systems. Besides the EHT there are other high resolution very long baseline interferometric (VLBI) telescopes such as the Global 3-mm VLBI array (GMVA) \footnote{https://www3.mpifr-bonn.mpg.de/div/vlbi/globalmm/} which operates at around 86 GHz. Imaging at resolutions of a few tens of microarcsec, these observatories will allow the imaging of the inner accretion region and the jet, in nearby accreting supermassive black holes.

Accretion processes around compact objects are the most energetic processes in the Universe and are responsible for powering the most luminous sources \citep{Novikov1973, Shakura1973, Lynden-Bell1974, Frank2002, Kato2008} in the sky such as Active Galactic Nuclei (AGN) \citep{Koratkar1999}, black hole binaries (BHBs) \citep{Remillard2006}, ultra-luminous X-ray sources \citep{Watarai2001} and similar objects. AGN can range from being superluminous ($L\sim 10^{45}$ erg/s) to having low luminosities ($L\sim 10^{40}$ erg/s). The most distant bright objects that we observe today are quasars which can be categorized as AGN. Many of the galaxies in the nearby universe host AGN such as the Seyferts, low luminosity nuclear emission-line regions (LINERs) and also Sgr A* in the center of our galaxy. 

Many AGN have accretion flows which are advection dominated. Such flows exist in sources where the accretion rate is much higher than that estimated from their bolometric luminosities. Advection dominated accretion flows (ADAFs) are also thought to exist in ultra luminous X-ray sources accreting at super-Eddington rates \citep{Katz1977, Begelman1979, Begelman1982, Abramowicz1988, Chen1993, Ohsuga2005,Abramowicz2013} which result in extreme high densities making the flow optically thick. Thus the heat, generated due to viscous dragging, is unable to escape and is then advected onto the black hole. On the other hand the accretion rate in most LINERs is sub-Eddington which results in an accretion flow with low densities and low optical depth \citep{Shapiro76, Ichimaru77, Rees82}. These low densities result in a two-temperature plasma. The excess heat generated, through viscous dragging in the heavy ions, cannot be efficiently radiated  by the electrons and is thus advected onto the black hole. Thus the ADAFs in sub-Eddington accretion flows, unlike the super-Eddington flows, are radiatively inefficient and are thus also known as radiatively inefficient accretion flows (RIAF).

There is a plethora of literature on sub-Eddington flows and RIAF models \citep{Ichimaru77, Rees82, Narayan94, Blandford99, Narayan2000, Quataert2000, Yuan2003} for low luminosity AGN (LLAGN). Recently a lot of work has also been done using GRMHD simulations in this area \citep{Davelaar2018,Monika2016, Monika2013, Monika2009} especially for M87 and Sgr A* but these simulations are computationally expensive and should be performed only once we have better constraints on the many parameters of the accretion flow of the system concerned. 

The accretion in LLAGN is accompanied by outflows, turbulence and strong magnetic fields which often lead to the formation of jets. The total emission from the flow is thus also affected by emission from jets and outflows \citep{Narayan94, Nemmen2007, Begelman2012}. It is important to consider all these effects while performing a multi-wavelength study of these LLAGN. Through this work we aim to show how a simple ADAF model with a combined jet model fit the observed high resolution data available in the literature to give us an idea about the possible physical processes responsible for the observed flux in different energy bands. The scientific community has previously used spectral energy distribution (SED) modeling/decomposition to probe the accretion physics \citep{Narayan94, Narayan95, Yuan2000, Yuan2003, Yuan05, Yuan14, Nemmen14, Xie2016, Li16}. With the models presented here, we first fit the high resolution (typically sub-arcsec in radio and arcsec scales in general) SED with an ADAF-Jet model and then use this to obtain the radial profile of emission from the accretion flow at micro-arcsec scales thus allowing us to  predict if the flow will be resolvable with the EHT, GMVA and the European VLBI network (EVN)\footnote{https://www.evlbi.org/home} plus RadioASTRON \footnote{Currently RadioASTRON is not functional but still we choose the 22 GHz band to have an idea about the emission profile in a range of frequency bands}. This then allows us to predict if the accretion flow in the sources of interest can be resolved by either the EHT or the GMVA given their resolution and RMS sensitivity. In addition to the dynamical model we also intend to infer if the emission from the accretion is affected by the presence of non-thermal electrons for our sample of LLAGN. We have selected sources for which we were able to get high resolution nuclear fluxes at different bands and for which there was recent observations in the X-rays  from the literature. 

This paper is organised as follows: In section [\ref{sec:Model}] we describe in detail the equations in sub-Eddington accretion flows and the effect of varying the various model parameters involved. We then discuss the jet model with its own set of model parameters and their effect on the spectrum. We subsequently describe the radiative processes involved in generating the spectrum. We then briefly describe our choice of the sample selected and the data used to constrain our model parameters in section [\ref{sec:Sample}]. In the following section [\ref{sec:Results}] we explain how the model parameters affect the total emission spectrum and the radial profile of emission at 86 GHz, which is the observing frequency of the GMVA. We then test the model with the SED of M87. In the same section we describe the fits to the data (given in the Appendix-\ref{sec:Data}) for our sample of LLAGN: Cen A, M84, Sombrero, NGC 3998 and NGC 4278. Also we obtain their radial flux profiles from the model fits to make a prediction about resolving the ADAF with the EHT (230 GHz) and the GMVA (86 GHz). Finally we provide the possible explanation of the various results obtained here through section [\ref{sec:Discussion}].

\begin{table*}
	\centering
	\caption{The observed parameter values and the references. The columns here correspond to the name of the sources, the logarithm of the ratio of the mass of the black hole to the solar mass, the distance in Mpc, the expected ring size (10.4 $R_g$) in $\mu$as, the Eddington ratio ($L_{bol}/L_{\rm Edd}$), the jet inclination angle and the references respectively. The references numbers mentioned in the last column are (1) \citet{Cappellari2009}, (2) \citet{Harris2010}, (3) \citet{Mezcua2014}, (4) \citet{Hada2013}, (5) \citet{Nemmen14} and (6) \citet{Giroletti2005} }
	\label{tab:Observables}
	\begin{tabular}{lcccccr} % four columns, alignment for each
		\hline
		Source & $log(M_{BH}/M_{\sun})$ & Distance & $\theta_{\rm Ring}$& Eddington Ratio & Inclination angle  & References\\
		       & 			&(Mpc)	   &	($\mu$as) & ($L_{Bol}/L_{\rm Edd}$) & $i$(\textdegree)& \\
		\hline
		NGC 5128 (Cen A) & $7.7$ & 3.8 & 1.5 & $5.0 \times 10^{-4}$ & $50 < i <80$ & 1,2,3 \\
		NGC 4374 (M84) & $8.9$ & 17.1 & 4.8 & $5.0 \times 10^{-6}$ & $30$ & 5 \\
 		NGC 4594 (Sombrero, M 104) & $8.5$ & 9.1 & 3.6 & $1.5 \times 10^{-6}$ & $25$ & 4,3,5.\\
 		NGC 3998 & $8.9$ & 13.1 & 6.2 & $1.0 \times 10^{-4}$ & $30$ & 5\\
 		NGC 4278 & $8.6$ & 14.9 & 2.7 &  $5.0 \times 10^{-6}$ & $2 < i <4$ & 5,6\\
		\hline
	\end{tabular}
\end{table*}

\section{Model Description}
\label{sec:Model}
The flux we obtain from the nucleus, with high resolution (ranging between arcsec to sub-arcsec scales) telescopes, may be composed of emission from the thermal and non-thermal electrons in the accretion region as well as the synchrotron emissions from the jet base. Thus in order to estimate the total flux, we need to have a clear picture of the various processes involved in the accretion region as well as in the jet. These along with the radiative transfer processes can result in the flux that we obtain in various bands. Hence in this section we present a simple 1 D model describing hot accretion flows, a jet and the various emission processes which finally result in generating the final spectrum.

\subsection{Accretion Model}
\begin{figure*}
	\includegraphics[width=2.5in,height=\columnwidth,angle=-90]{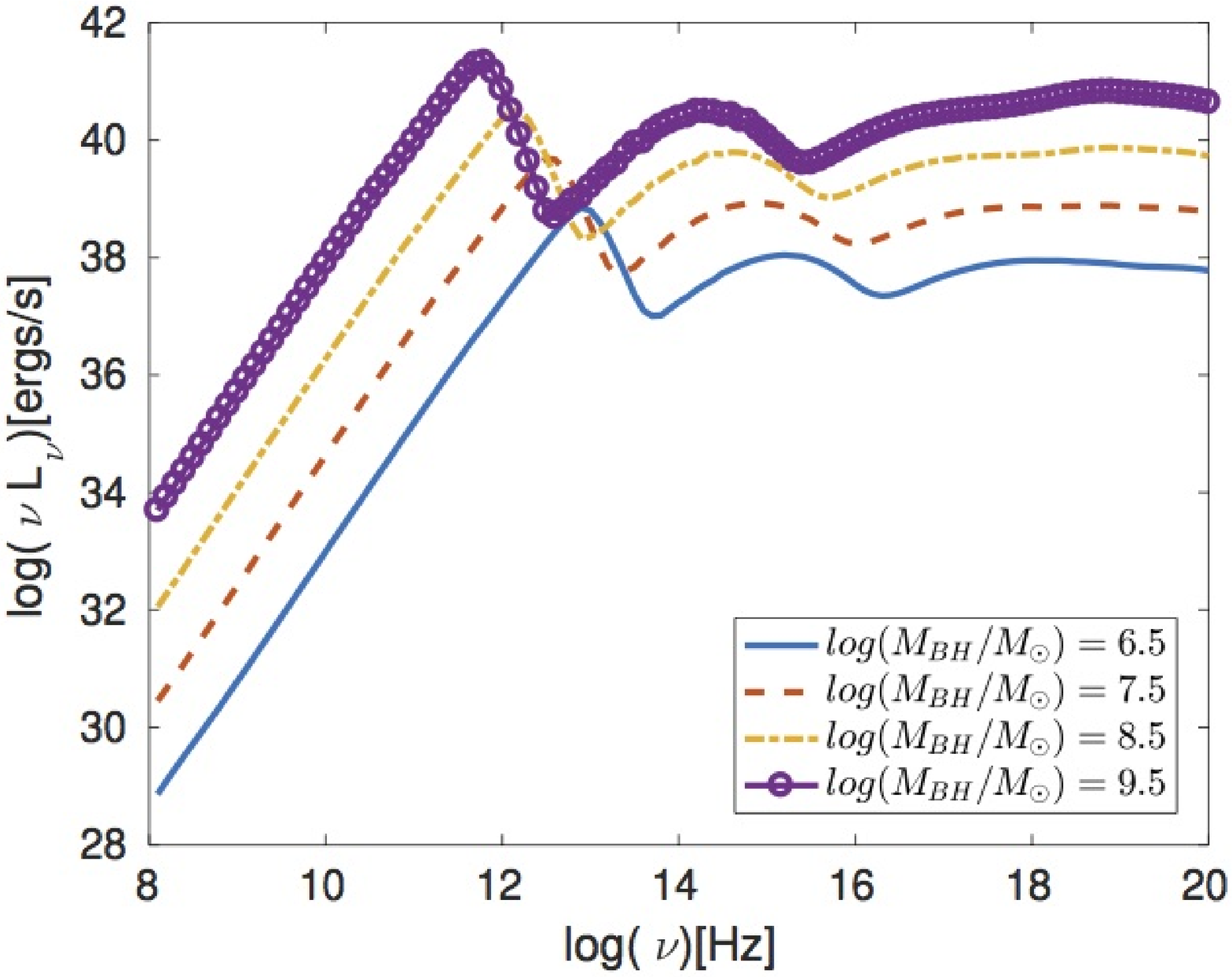}
	\includegraphics[width=2.5in,height=\columnwidth,angle=-90]{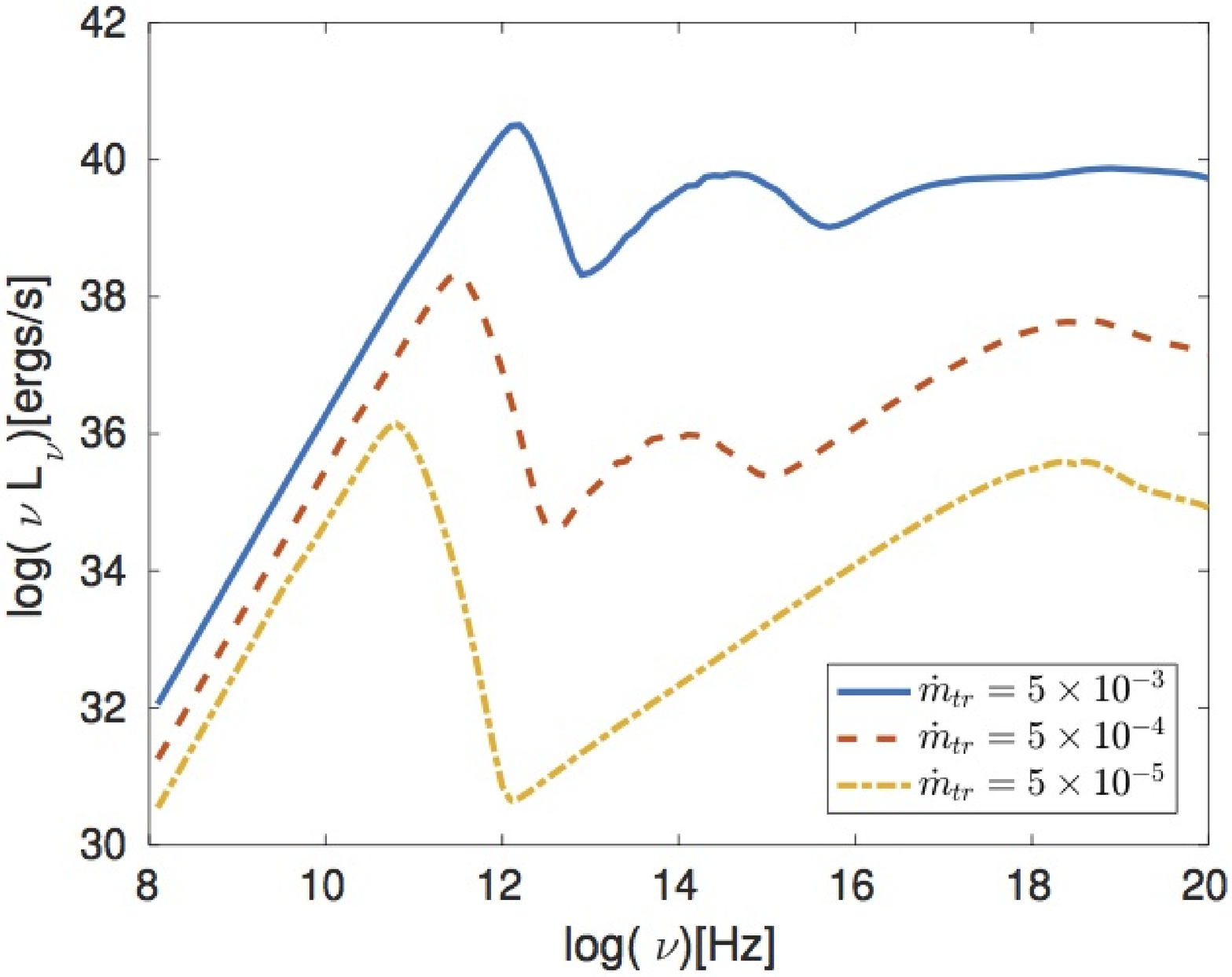} \\
	\includegraphics[width=2.5in,height=\columnwidth,angle=-90]{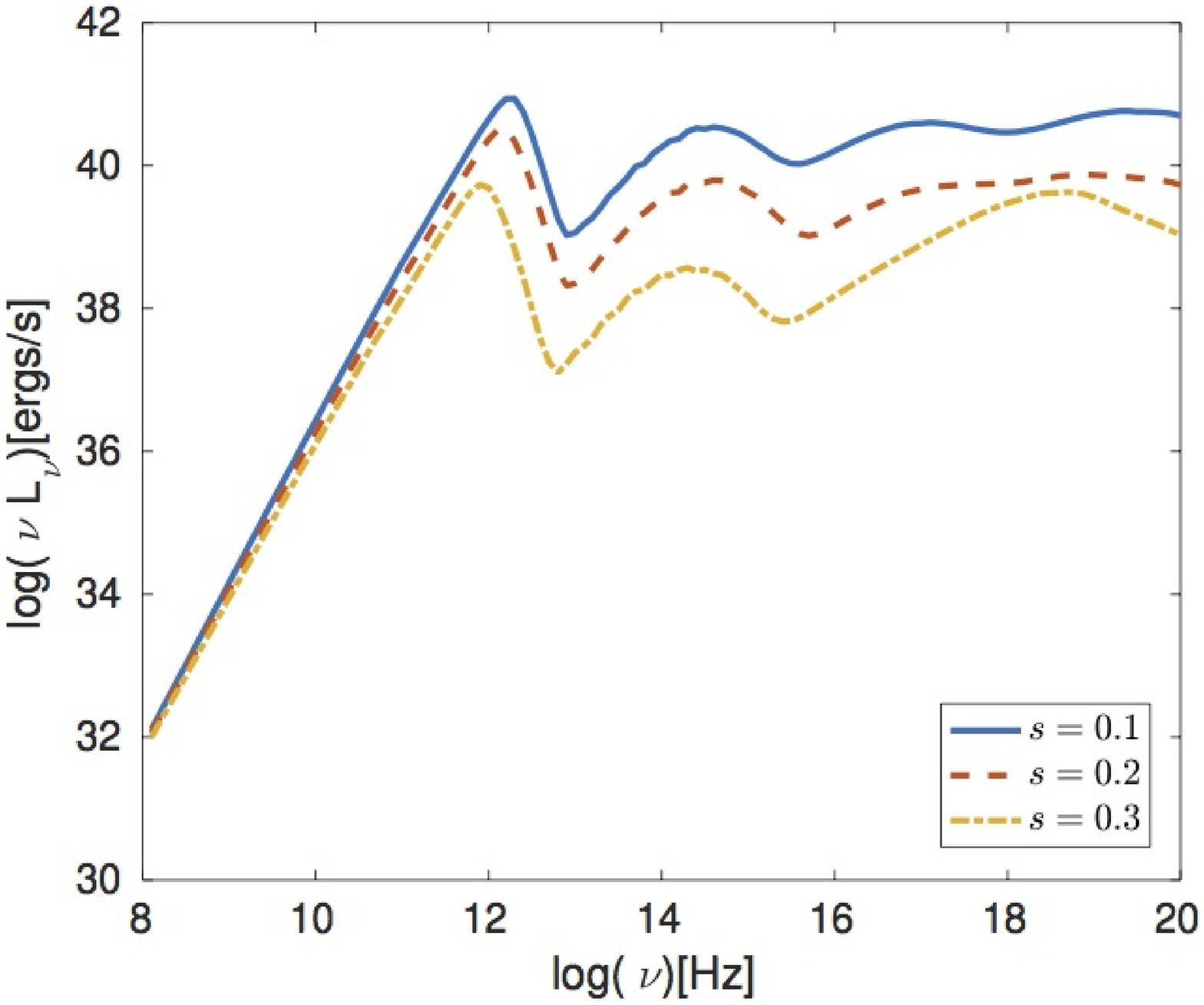}
	\includegraphics[width=2.5in,height=\columnwidth,angle=-90]{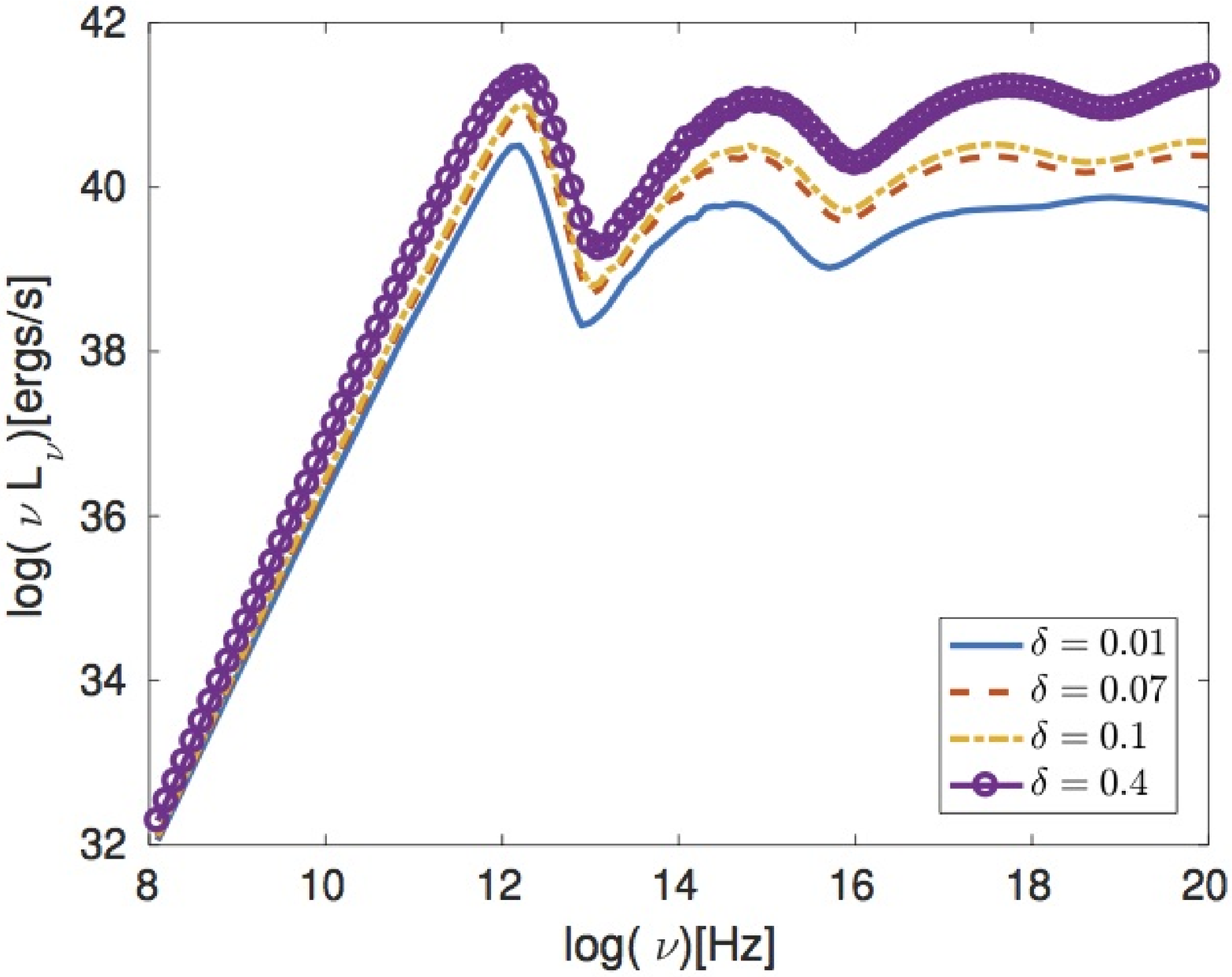}
   \caption{The effect of changing individual parameters of the ADAF model on the predicted SED of the full accretion flow. {\it Top Left Panel}: Spectral variation with varying black hole masses. {\it Top Right Panel}: Spectral variation with varying accretion rates {\it i.e.} $\dot{m}_{tr}$. {\it Bottom Left Panel}:  Spectral variation with variation in the outflow parameter ($s$). {\it Bottom Right Panel}: Spectral variation with variation in the energy injection parameter ($\delta$). To obtain the ADAF SED, we have assumed the following ADAF model parameters in general: $log (M_{\rm BH}/M_{\odot})=8.5$, $\alpha =0.3$, $\beta = 0.9$, $\delta = 0.01$, $\dot{m}_{\rm tr}=5 \times 10^{-3}$, $R_{tr} = 10^4 R_g$ and $s = 0.2$.} 
    \label{fig:complots}
\end{figure*}

\begin{figure*}
	\includegraphics[width=2.5in,height=\columnwidth,angle=-90]{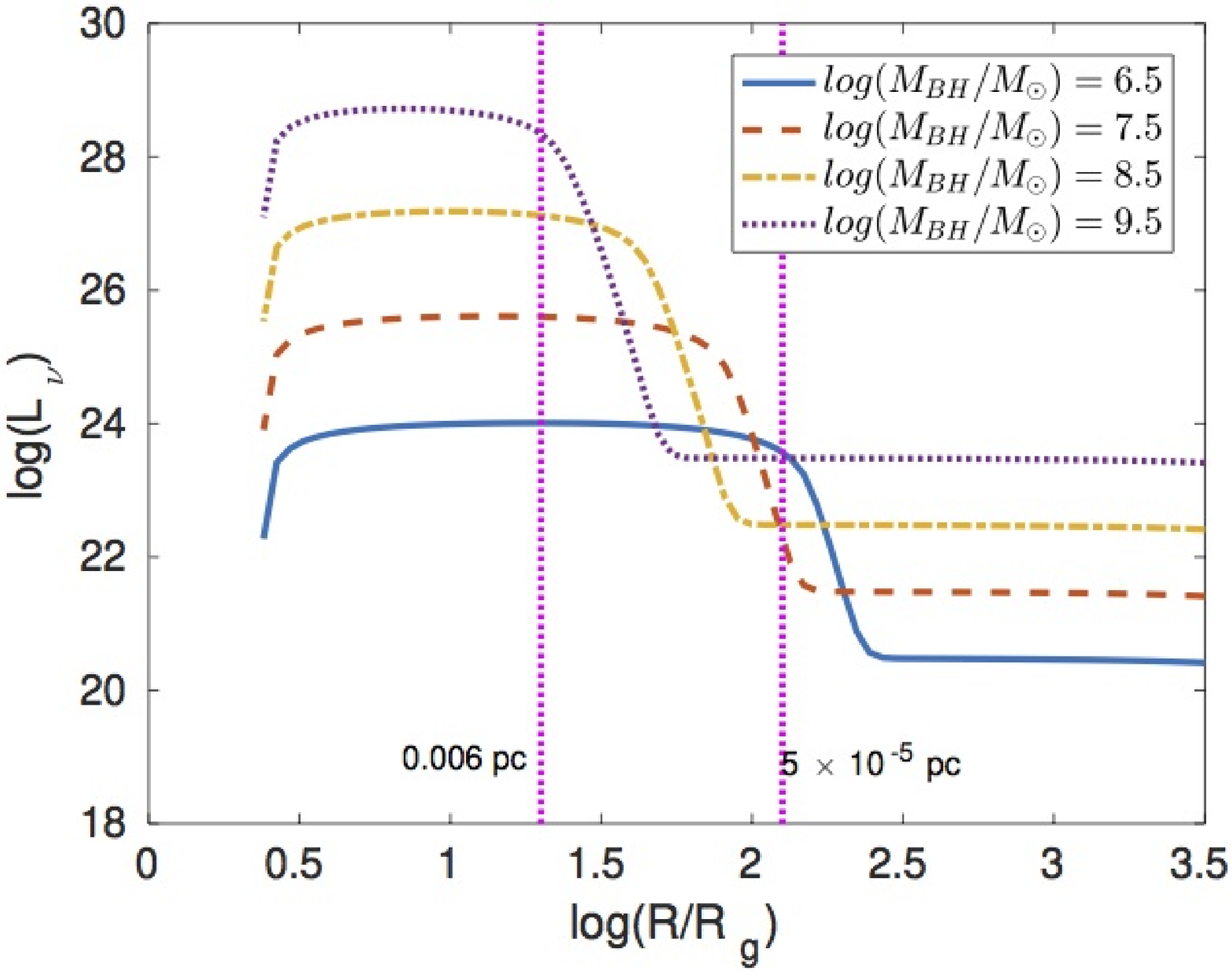}
	\includegraphics[width=2.5in,height=\columnwidth,angle=-90]{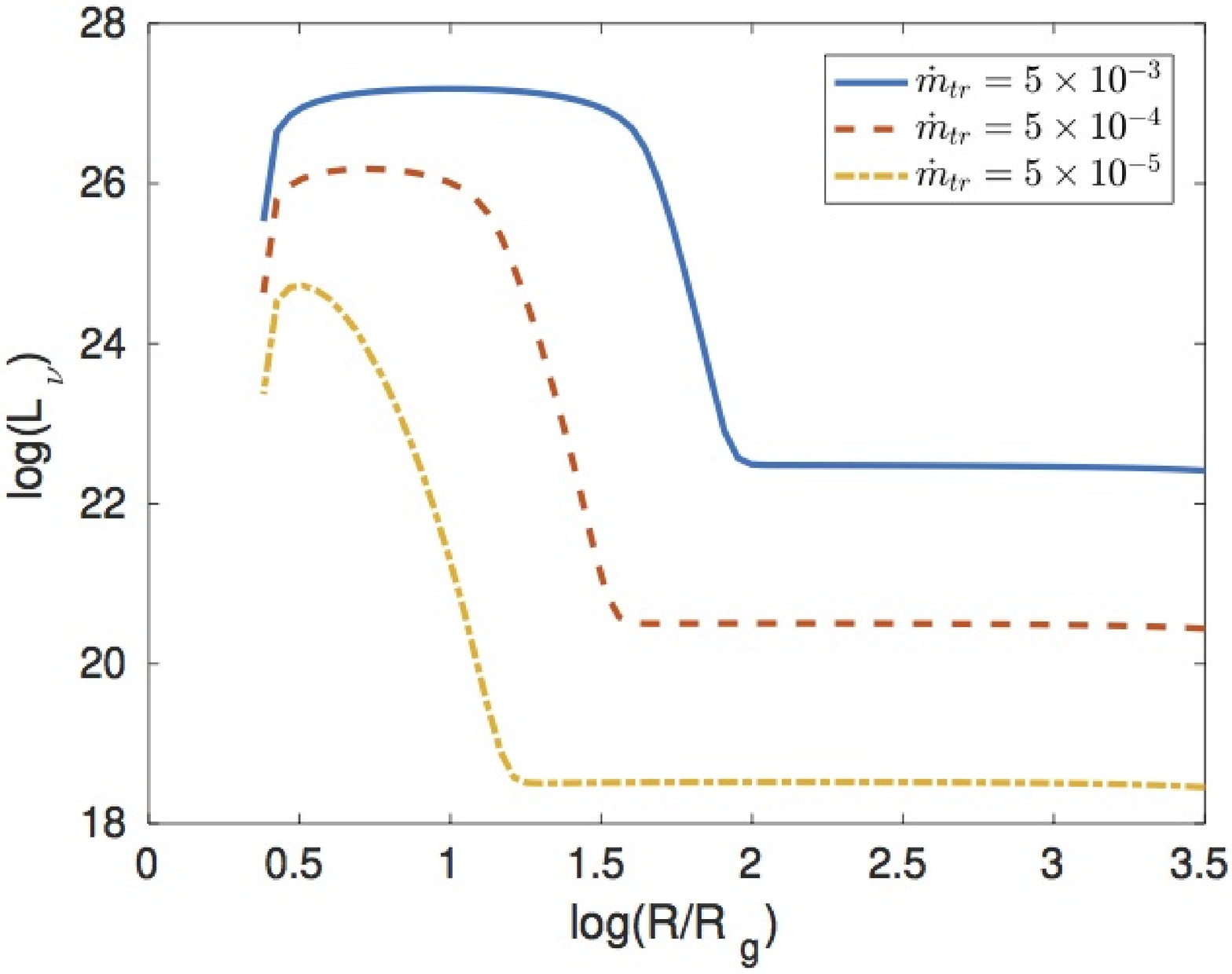} \\
	\includegraphics[width=2.5in,height=\columnwidth,angle=-90]{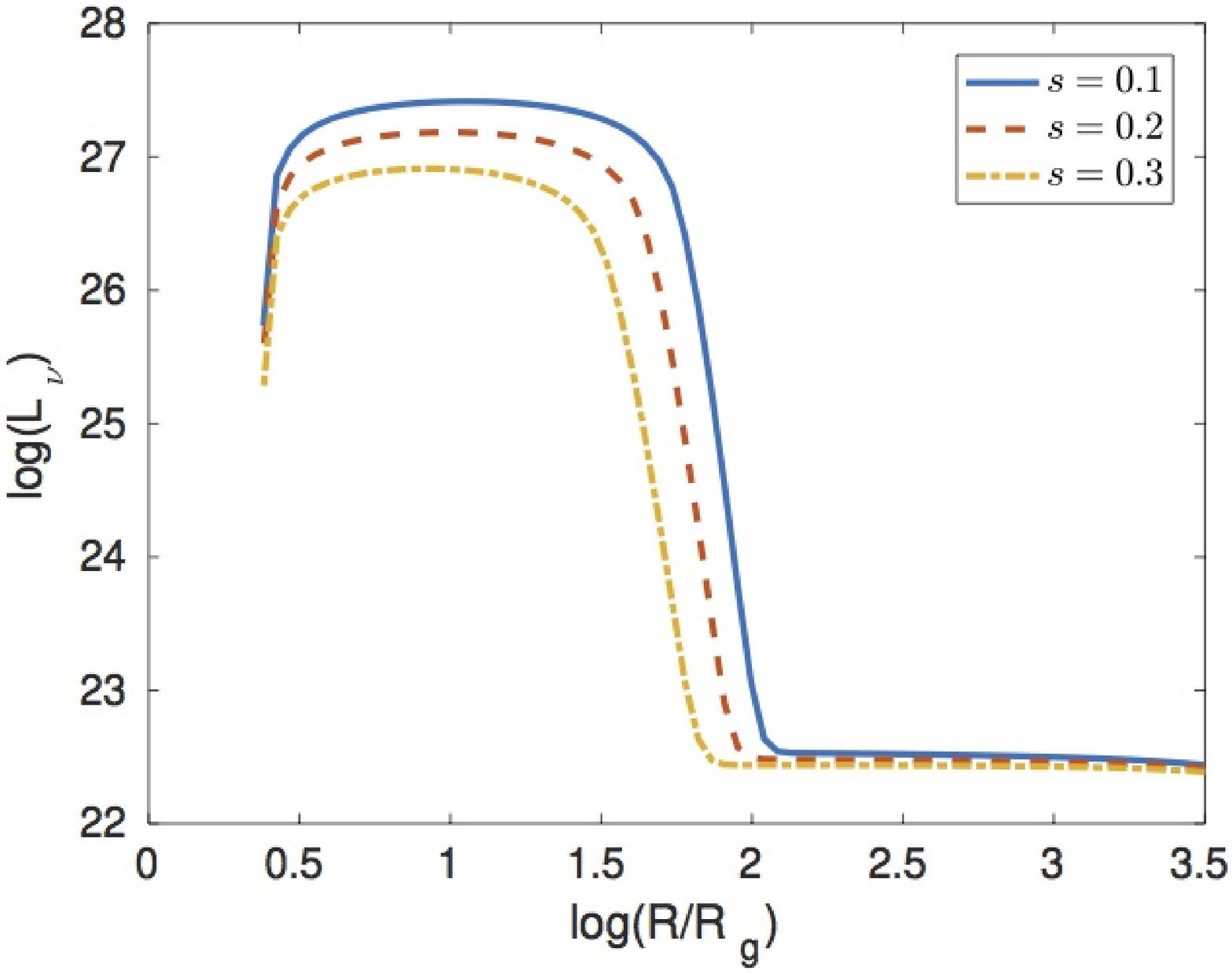}
	\includegraphics[width=2.5in,height=\columnwidth,angle=-90]{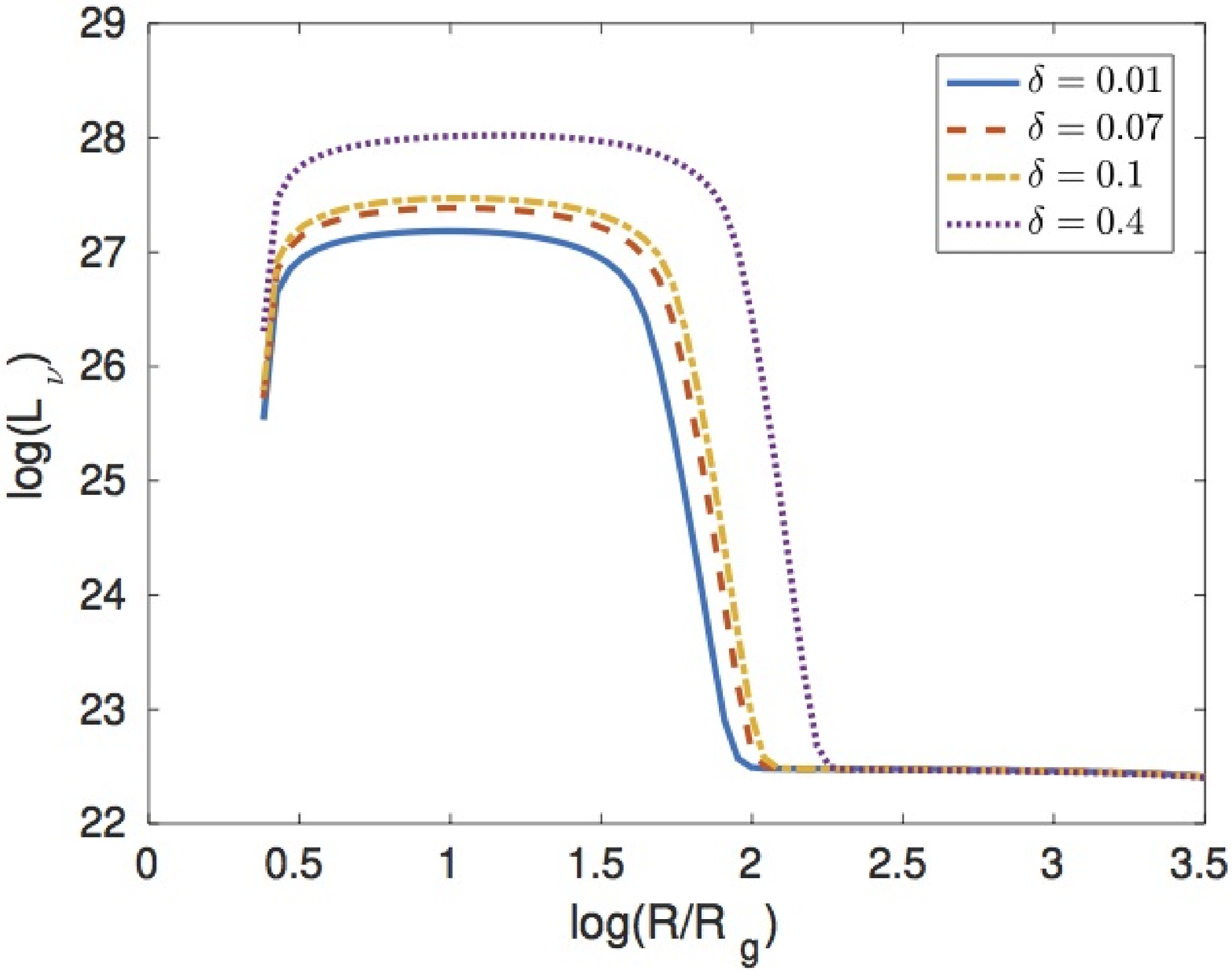}
   \caption{Variation in the radial emission profiles at 86 GHz with the same parameter variations shown in Fig.[\ref{fig:complots}]. In addition, in the top left panel we have marked the knee of the emission profile with magenta dotted vertical lines for the two extreme mass limits, in order to display their actual physical extent.} 
    \label{fig:complotsradial}
\end{figure*}

 % used for referring to this section from elsewhere
Most of the emission from the nuclear region of a LLAGN originates from the accretion flow very close to the supermassive black hole sitting at the center. At regions in the extreme proximity of the black hole, the emission and accretion processes will be greatly affected by GR effects but at distances close to 10 times the Schwarzschild radius an accretion model with a weak GR effect can be a good approximation. They can be modeled using a sub-Eddington accretion flow \citep{Shapiro76, Ichimaru77, Rees82} where the accretion rate is much smaller than the Eddington rate and the gas reaches its virial temperature. Owing to such small accretion rates, the flow has very low densities and is optically thin. The excess heat, which is generated due to viscous dragging, is unable to escape via radiation by electrons. This results in a two temperature plasma with ions at a temperature higher than the electrons. Owing to the radiative inefficiency of the heavy ions, the excess heat is thus advected onto the black hole. Due to the various processes involved, these flows are also called {\it hot accretion flows}, {\it radiatively inefficient accretion flows (RIAF)} or {\it advection dominated accretion flows (ADAF)}.  The excess temperature leads to an expansion of the ADAF, thus making it geometrically thick. These disks are accompanied by outflows, which implies that the accretion rate is not constant along the radial accretion flow. We investigate the evolution of the dynamical equations in an ADAF model tailored to LLAGN \citep{Yuan05}. The following dynamical equations \citep{Yuan14} can be set from the laws of conservation of mass, radial momentum, angular momentum and energy :

\begin{eqnarray}
  \dot{M}(R)=\dot{M}({R_{tr}})&\left(\frac{R}{R_{\mathrm{tr}}}\right)^s&=4\pi \rho R H |v|. \label{mascon} \\
%\end{equation}
%
%\begin{equation}
 v\frac{dv}{dR}-\Omega^2 R = -\Omega_K^2 R &-& \frac{1}{\rho}\frac{d}{dR}(\rho c_s^2). \label{radmom}\\
% \end{equation}
% 
% \begin{equation}
 \frac{d\Omega}{dR} &=& \frac{v\Omega_K (\Omega R^2 - j)}{\alpha R^2 c_s^2}. \label{angmom}\\
% \end{equation}
%  
%  \begin{eqnarray}
  \rho v\left(\frac{de_i}{dR}-\frac{p_i}{\rho^2}\frac{d\rho}{dR}\right) &=& (1-\delta)q^+ - q^{ie}. \nonumber \\
   \rho v\left(\frac{de_e}{dR}-\frac{p_e}{\rho^2}\frac{d\rho}{dR}\right) &=& \delta q^+ + q^{ie} - q^- . \label{enercon}
\end{eqnarray}
Here $R_{\rm tr}$ is the truncation radius, $s$ is the parameter that quantifies outflow from accretion, $H=c_s/\Omega_K$ is the scale height, $\rho$ is the gas density, $\Omega$ is the angular momentum of the in-falling gas, $\Omega_K$ is the angular momentum of the Keplerian orbit, $v$ is the radial velocity of the gas, $c_s$ is the sound speed, $j$ is the angular momentum at the gravitational radius $R_g$ and is an eigen value for the system under consideration, $\alpha$ is the viscosity parameter, $e_e$ and $e_i$ are the specific internal energies of the electrons and ions, respectively, $\delta$ is the fraction of the viscous energy ($q^+$) that goes into heating the electrons \citep{Xie12, Chael18}, $q^{ie}$ the energy that is exchanged between electrons and ions and $q^{-}$ is the energy lost via radiation. It should be noted that eq.[\ref{mascon}] includes the case of outflows while eq.[\ref{enercon}] is the modified energy conservation equation for two temperature plasmas. These equations are solved simultaneously using proper boundary conditions \citep{Yuan2000} at the truncation radius, sonic radius and at $R_g$. In this work, we express the mass accretion rate $\dot{M}$ in terms of the Eddingtion accretion rate $\dot{M}_{\rm Edd}$ through the dimensionless parameter $\dot{m}$ as $\dot{M}(R)=\dot{m}(R)\dot{M}_{\rm Edd}$. This parameter at $R_g$ is equal to the Eddington ratio ($L_{\rm Bol}/L_{\rm Edd}$) in case of thin disk accretion flows but is higher for ADAF. We use the Eddington ratio as a lower limit for $\dot{m} (R_g)$. We vary this parameter by varying $\dot{m}(R_{\rm tr})$ (from now on $\dot{m}_{\rm tr}$) and $s$ using eq.[\ref{mascon}]. The pressure ($p_i$ and $p_e$) in eq.[\ref{enercon}] is the gas pressure ($p_{\rm gas}=p_i +p_e$) expressed in terms of the total pressure ($p_{\rm tot}=p_{\rm gas}+p_{\rm magnetic}$) as $p_{gas}=\beta p_{\rm tot}$. Variations in $\alpha$ are accompanied by similar variations in $\beta$ \citep{Yuan14}. Thus for all our sources we fix $\beta=0.9$ and the viscosity parameter $\alpha=0.3$ as there is not much variation in the modeled SED on varying these parameter values. Since $R_{\rm tr}$ is set as an initial condition, we use $R_{\rm tr}= 10^{4}R_g$. The value of $s$ can vary between 0 and 1. For RIAFs like our sources, $\delta$ varies between 0.01 and 0.5. Electrons in the flow emit over a large range of frequencies from radio to gamma rays via synchrotron, bremsstrahlung and inverse Compton scattering. We have shown in Fig.[\ref{fig:complots}] and Fig.[\ref{fig:complotsradial}] how the variation of the parameters affect the total spectrum and the radial profile of emission at 86 GHz (for GMVA) respectively. 

\begin{figure*}
	\includegraphics[width=2.5in,height=\columnwidth,angle=-90]{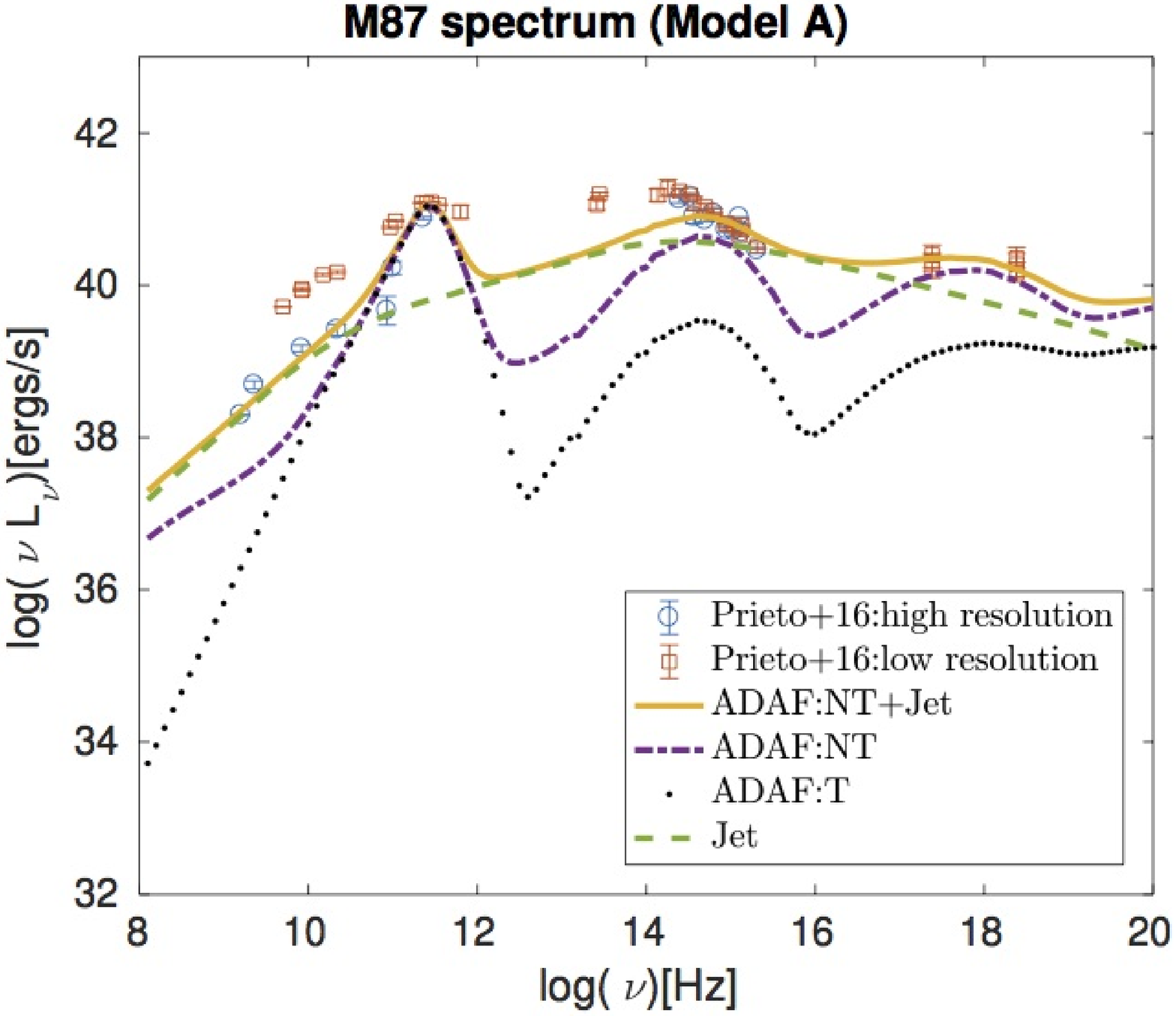}
	\includegraphics[width=2.5in,height=\columnwidth,angle=-90]{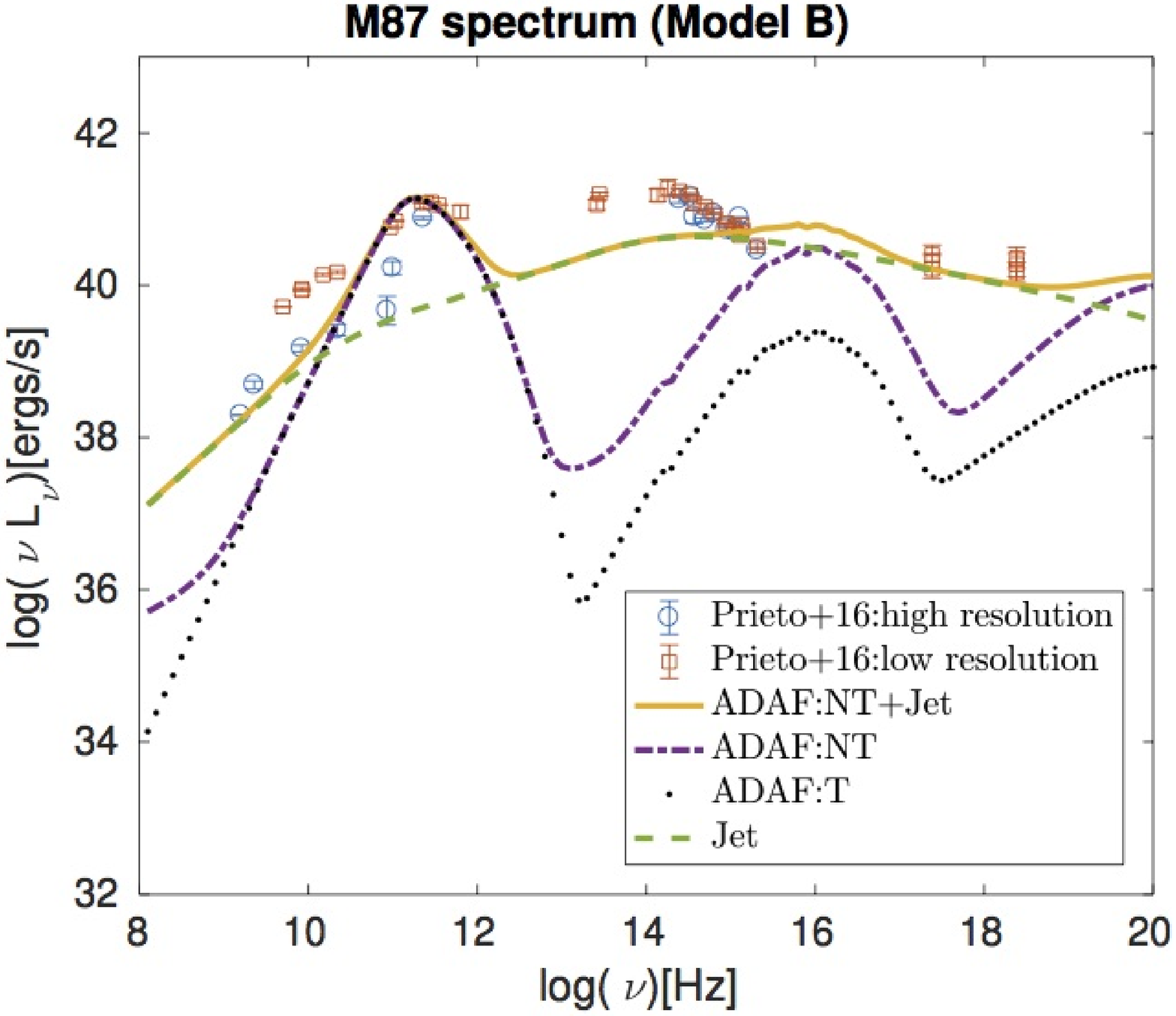}\\
	\includegraphics[width=2.5in,height=\columnwidth,angle=-90]{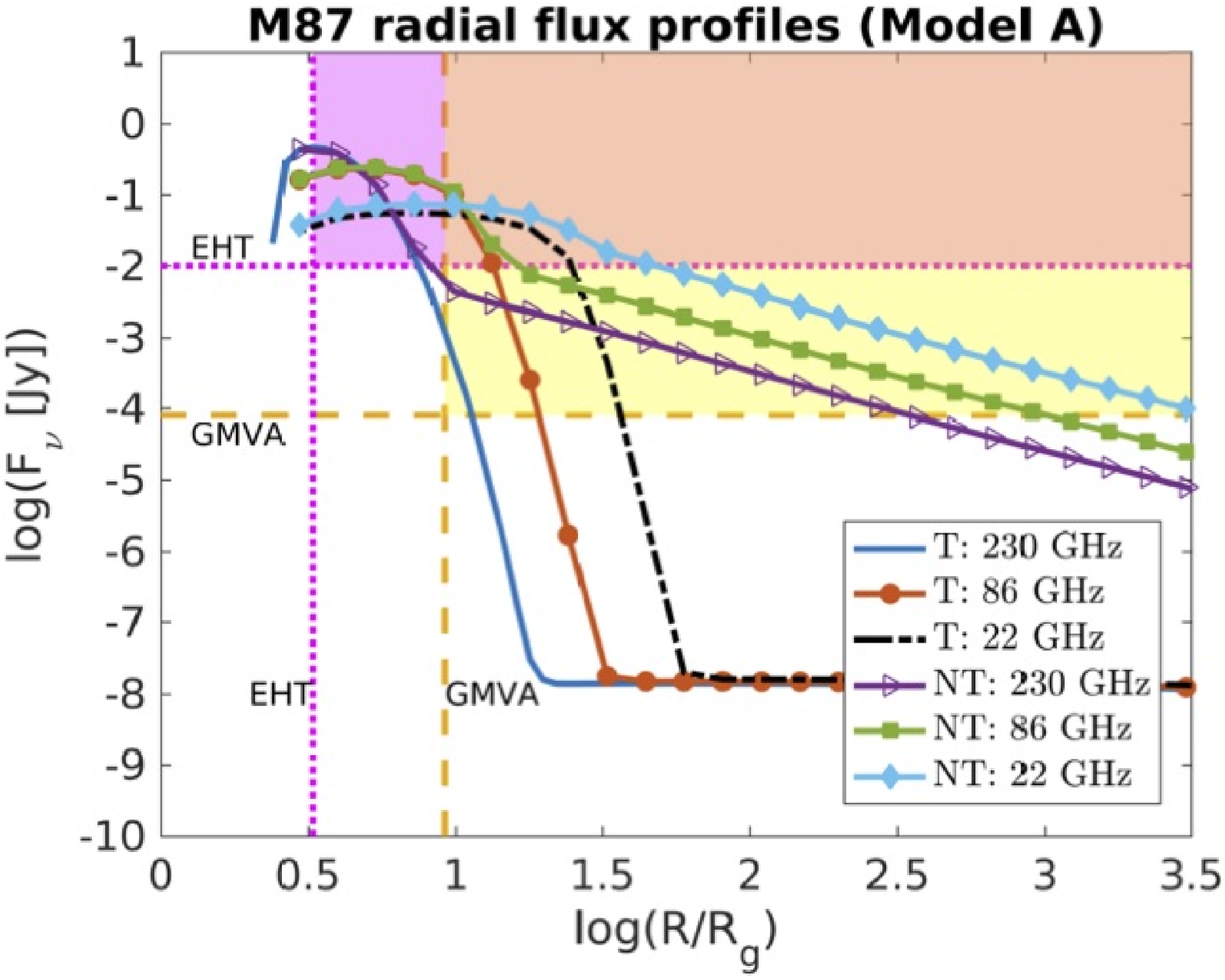}
	\includegraphics[width=2.5in,height=\columnwidth,angle=-90]{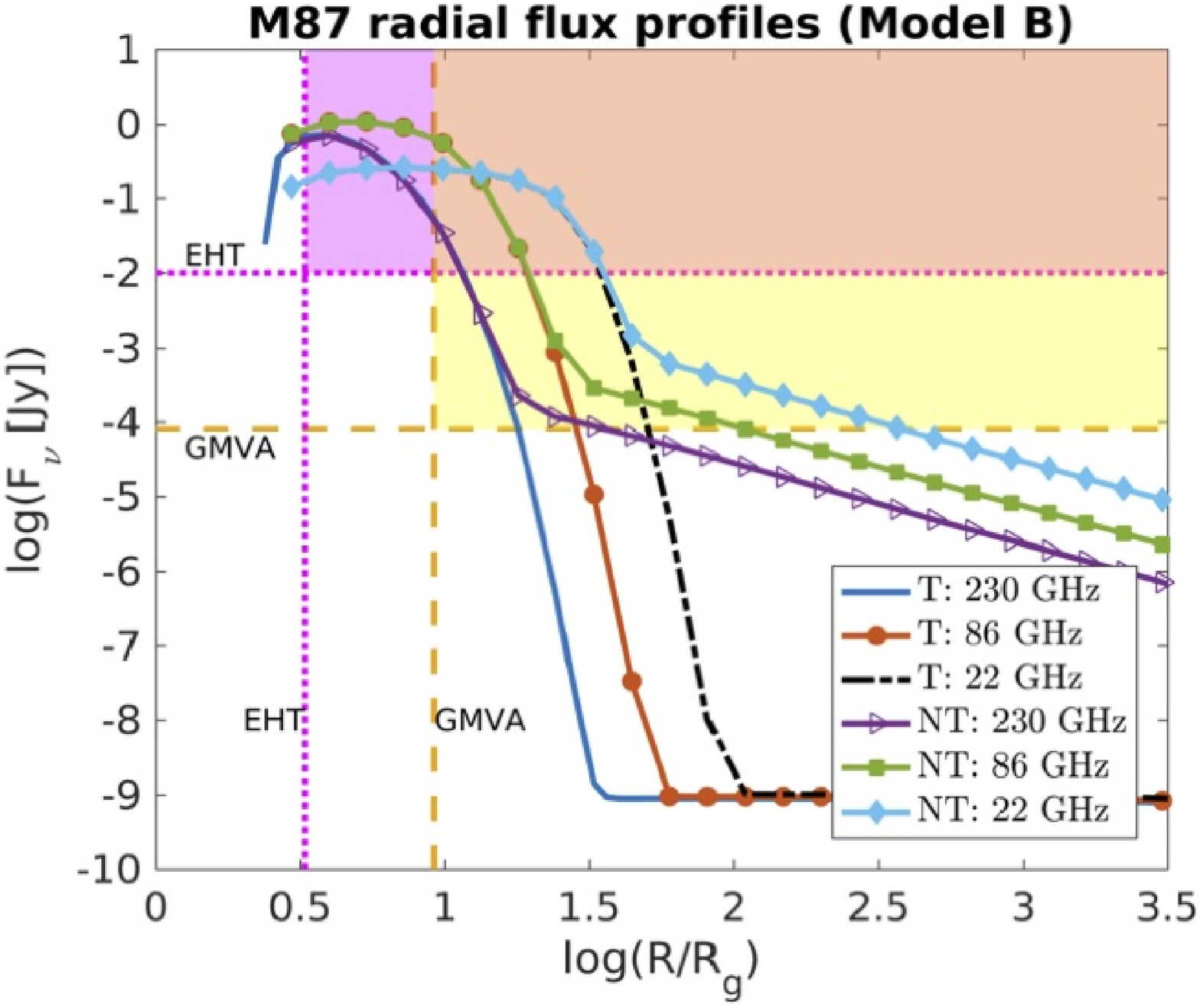}
   \caption{{\it Top Panels}: The multi-wavelength data and fitted model spectra for M87 with the left and right panel corresponding to model A and model B (see Table - [\ref{TestM87}] ) respectively. The blue circles and red squares correspond to high ($\sim$ milli-arcsec) and low ($\sim 0.4$ arcsec) resolution observations respectively. Fitted SEDs are shown with black dotted lines (emission only from the thermal electrons of the ADAF model), purple dotted dashed lines (emission from thermal plus non-thermal electrons of the ADAF), green dashed lines (emission from the Jet) and the yellow solid lines (emission from the jet and ADAF model with thermal plus non-thermal electrons). {\it Bottom Panels}: The radial flux profiles of the ADAF for model A and model B are shown on the left and right panels respectively. The pink dotted vertical and horizontal lines correspond to the 25 $\mu$as resolution and the expected 10 mJy rms sensitivity for an 8 hour integration time of the EHT. The yellow dashed vertical and horizontal lines correspond to the 70 $\mu$as resolution and the expected 0.08 mJy rms sensitivity for an 8 hour integration time of the GMVA. The radial flux profiles at 230 GHz (EHT) are shown with the blue solid (emission from the thermal electrons) and purple arrowed solid lines (emission from thermal plus non-thermal electrons). Similarly the profile at 86 GHz (GMVA) with red circular solid (thermal) and green squared solid lines (thermal plus non-thermal) and the ones at 22 GHz (EVN) with the black dotted dashed (thermal) and light blue diamond solid lines (thermal plus non-thermal). The pink shaded area and the yellow shaded area correspond to the region of observability of the EHT and GMVA, respectively, for an 8 hour integration time. The orange shaded region corresponds to the common area of observability of both the telescopes.}
    \label{fig:M87}
\end{figure*}

\begin{table*}\label{TestM87}
% 	\centering
	\caption{ The best fit ADAF+Jet model parameter values for the two models, A and B, for M87. The columns represent the mass outflow rate at the truncation radius ($\dot{m}_{\rm tr}$) in terms of Eddington rate $\dot{M}_{\rm Edd}$, the energy injection parameter $\delta$, outflow parameter $s$, the angular momentum parameter ($j$) at the gravitational radius, power law index ($p_l$) of electrons in the ADAF, the fraction ($\eta$) of electrons in the disk which are boosted to power law, the jet outflow rate ($\dot{m}_{\rm jet}$) as a function of $\dot{M}_{\rm Edd}$, the power law index ($p_{\rm jet}$) of electrons in the jet, the energy input parameters for electrons ($\epsilon_e$) and magnetic field ($\epsilon_{B}$) and finally the fraction of electrons ($\xi$) in the jet with power law distribution.}
	\label{tab:Parameter}
	\begin{tabular}{lcccccccccccr} % four columns, alignment for each
		\hline
		Model  & $\dot{m}_{tr}$ & $\delta$ & $s$ & $j$ & $p_l$ & $\eta$& $\dot{m}_{jet}$ & $p_{jet}$ & $\epsilon_e$ & $\epsilon_b$ & $\xi$\\
		\hline
		Model A & $4.2 \times 10^{-4}$ & $0.1$ & $0.1$ & $0.7999$ & $3.0$ & $0.015$ & $1.0 \times 10^{-8}$ & $2.6$ & $0.0009$ & $0.0006$ & $0.01$\\
		Model B & $1.2 \times 10^{-4}$ & $0.5$ & $0.3$ & $1.8360$ & $3.0$ & $0.015$ & $1.0 \times 10^{-8}$ & $2.5$ & $0.0009$ & $0.0006$ & $0.01$\\
		\hline
	\end{tabular}
\end{table*}

\subsection{The Jet}
The accretion dynamics is more complex due to turbulence, the presence of magnetic fields, hot spots and outflows. \citet{Narayan94, Narayan95, Blandford99} postulate that hot accretion flows should have strong winds followed by the formation of jets. This is supported by observational evidence which suggests that almost all LLAGN are radio-loud \citep{Falcke00, Nagar00, Ho02}.  The jet dynamics is generally assumed to arise from a combination of magnetic fields and rotation. The most accepted theoretical models are the Blandford-Znajek (BZ) model, \citep{Blandford77} which states that the primary source of energy in the jet is the rotational energy of the black hole, while the Blandford-Payne (BP) model \citep{Blandford82} suggests that it is due to the rotational energy of the accretion flow. Independent of the origin of the jet, it is often necessary to include a jet to explain the observed SED of most LLAGN \citep{Nemmen14, Li16}. The most powerful jets can be produced by highly-magnetized versions of ADAF, i.e. the magnetically arrested disk (MAD, \citet{Tchekhovskoy, Chael2019}). In this work, we use a phenomenological model \citep{Spada01,Yuan05,Xie2014} to describe the jet, which is sufficient to model the SED. It is assumed to be composed of normal plasma, consisting of electrons and protons, with velocities determined by a bulk Lorentz factor $\Gamma_j=10$ (typical for jets in AGN as in \citet{Lister2016}).

A fraction $\xi$ of electrons is boosted to a power law (power law index $p_{jet}$) energy distribution due to internal shocks within the jet. The accelerated electrons, in the high energy part of the power law spectrum, cool down due to radiative cooling. Thus they now acquire a distribution index of $1+p_{jet}$ \citep{Rybicki79}. Parameters defining the fraction of the shock energy that goes into electrons and magnetic fields, $\epsilon_e$ and $\epsilon_B$ respectively, are included. These micro physical parameters are taken to be constant along the jet direction making a simplified but reasonable assumption.

Given the above model, the synchrotron emission from the boosted electrons can now be evaluated. Inverse Compton scattering of these synchrotron photons is almost negligible owing to the small scattering optical depth (but see \citet{Markoff2005}). Thus estimating the SED of the jet is moderately simple. The high-energy part (e.g. UV and X-ray bands) of the SED is a power law. Since the electrons responsible for the X-ray radiation are cooled, the photon index is then $1+p_{\rm jet}/2$ \citep{Rybicki79}. The low-energy part, from radio up to IR, is also a power law. The spectrum is flat or slightly inverted with spectral index $\alpha_{\rm jet}\approx0-0.5$ (where $\alpha_{\rm jet}$ is defined through $F_{\nu}\propto \nu^{\alpha_{\rm jet}}$) because of self-absorption. The power law index $p_{\rm jet}$ is related to $\alpha_{\rm jet}$ using the relation $1-\alpha_{\rm jet}=1+p_{\rm jet}/2$. In general there is a degeneracy in the parameters for the model. The mass loss rate $\dot{M}_{\rm jet}$ is sensitively coupled with the jet outflow velocity $V_{\rm jet}$ (assumed constant for all our LLAGN) which controls the beaming effect and gas density. The radiation at every frequency band is proportional to the parameters $\dot{M}_{\rm jet}$, $\epsilon_e$ and $\epsilon_B$. While the radiation at high frequencies (e.g. X-ray and UV) is more sensitive to $\epsilon_e$ compared to that at low frequencies (e.g. radio and IR), $\epsilon_B$ and also $\dot{M}_{\rm jet}$ (with weaker effects) show opposite effects ({\it i.e.} more sensitive at lower frequency bands). The role of $\xi$ is more complex. Enhancing its value ({\it i.e.} reducing the mean energy of the power-law electrons in the jet) will reduce the X-ray radiation but enhance the radio emission. However, the X-ray spectral shape is primarily determined by $p_{\rm jet}$ and also affected by $\xi$ and $\epsilon_e$. The variability of the jet emission, with the various parameter values, has been shown by \citet{Xie2014}.

\begin{table*}
% 	\centering
	\caption{The ADAF model parameter values that fit to the data. The model TE corresponds to emission only from the thermal electrons in the flow while model PL corresponds to emission from the power law electrons as well. The other columns represent the mass outflow rate at the truncation radius ($\dot{m}_{\rm tr}$) in terms of Eddington rate $\dot{M}_{\rm Edd}$, the energy injection parameter $\delta$, outflow parameter $s$, the angular momentum parameter ($j$) at the gravitational radius, power law index ($p_l$) of electrons in the ADAF, the fraction ($\eta$) of electrons in the disk which are boosted to power law, the jet outflow rate ($\dot{m}_{\rm jet}$) as a function of $\dot{M}_{\rm Edd}$, the power law index ($p_{\rm jet}$) of electrons in the jet, the energy input parameters for electrons ($\epsilon_e$) and magnetic field ($\epsilon_{B}$) and finally the fraction of electrons ($\xi$) in the jet with power law distribution.}
	\label{tab:Parameter}
	\begin{tabular}{lccccccccccccr} % four columns, alignment for each
		\hline
		Source & model & $\dot{m}_{tr}$ & $\delta$ & $s$ & $j$ & $p_l$ & $\eta$& $\dot{m}_{jet}$ & $p_{jet}$ & $\epsilon_e$ & $\epsilon_b$ & $\xi$\\
		\hline
		Cen A & PL & $2 \times 10^{-2}$ & $0.2$ & $0.3$ & $1.6455$ & $3.5$ & $0.015$ & $5.5 \times 10^{-5}$ & $2.1$ & $0.001$ & $0.0007$ & $0.01$\\
		M84 & TE & $2.9 \times 10^{-3}$ & $0.09$ & $0.3$ & $1.4089$ & - & -& $4.0 \times 10^{-7}$ & $2.6$ & $0.0009$ & $0.0006$ & $0.01$\\
 		NGC 4594 & TE & $5 \times 10^{-3}$ & $0.01$ & $0.2$ & $1.1098$ & - & -& $9.0 \times 10^{-7}$ & $2.4$ & $0.0007$ & $0.0001$ & $0.01$\\
 		NGC 3998 & TE & $7 \times 10^{-3}$ & $0.2$ & $0.3$ & $1.5533$ & - & -& $1.0 \times 10^{-6}$ & $2.1$ & $0.0005$ & $0.0001$ & $0.01$\\
 		NGC 4278 & PL & $2 \times 10^{-3}$ & $0.08$ & $0.3$ & $1.3958$ & $3.1$ & $0.07$& $3.0 \times 10^{-9}$ & $3.0$ & $0.0001$ & $0.015$ & $0.05$\\
		\hline
	\end{tabular}
\end{table*}

\subsection{The Spectral Energy Distribution (SED) and emission profiles}
The important parameters that determine the SED from the ADAF disk are the temperature, density of electrons and the velocity profiles of the gas that we obtain as the solution to the dynamical equations mentioned above. Assuming the disk is isothermal in the vertical direction, the spectrum of unscattered photons at a given radius is calculated by solving the radiative transfer equation in the vertical direction of the disk based on the two-stream approximation \citep{Rybicki79}. Since the gas close to the black hole is hot, optically thin and magnetized, the processes which significantly contribute to the emission are synchrotron and bremsstrahlung \citep{Manmoto97}. The presence of electrons comptonizes \citep{Coppi90} these photons to modify the total SED. 

Processes such as magnetic reconnection, weak shocks and turbulent dissipation can accelerate a fraction of the thermal electrons to a non-thermal power-law distribution, which also emit via synchrotron emission \citep{Yuan2003,Yuan05,Liu2013}. The power-law electrons in the jet lead to an enhanced contribution of the synchrotron emission. In our model we distinguish the two cases when the emission from the non-thermal electrons affect the emission and the one where the emission is just from the thermal electrons. The process of emission by the power-law or non-thermal electrons from the ADAF is included following the method of \citet{Ozel} where the important parameters are the power-law index $p_l$ and the fraction of thermal electrons $\eta$ boosted to power-law electrons. It can be often difficult to distinguish the contribution of non-thermal electrons in the jet to that from non-thermal electrons in the ADAF in case of strong jetted systems at low frequency radio bands. With upcoming precise high frequency observations and spatially resolved observations, we expect that it will be possible to distinguish these components.

The emission in the radio is primarily due to synchrotron emission. At lower frequencies there is synchrotron self absorption within the ADAF. The presence of non-thermal electrons can modify the emission a little at lower frequencies but the effect of self-absorption can still be seen. Only synchrotron emission from the jet can result in higher fluxes at low frequencies. On the other hand the spectrum at higher frequencies (X-ray) is primarily determined by the thermal bremsstrahlung emission. The comptonization of both these emissions then modifies the entire spectrum resulting in peaks in infrared, optical and also in the X-ray band depending on the model parameters.  Our model can constrain the high energy regions only up to 0.5 MeV ($10^{20}$ Hz). Energies above this are affected by hadronic processes \citep{Niedzwiecki2013} and are also sensitive to the black hole spin, which determine the high ion temperature close to the black hole. Both these effects are beyond the scope of this work. %We use the example of M87 for which we consider both high and low resolution data from \citet{Prieto2016} and describe the the importance of the emission of power-law electrons in the ADAF to obtain a better fit to the data as shown in Fig.[\ref{fig:M87}].

\begin{figure*}
	\includegraphics[width=2.5in,height=\columnwidth,angle=-90]{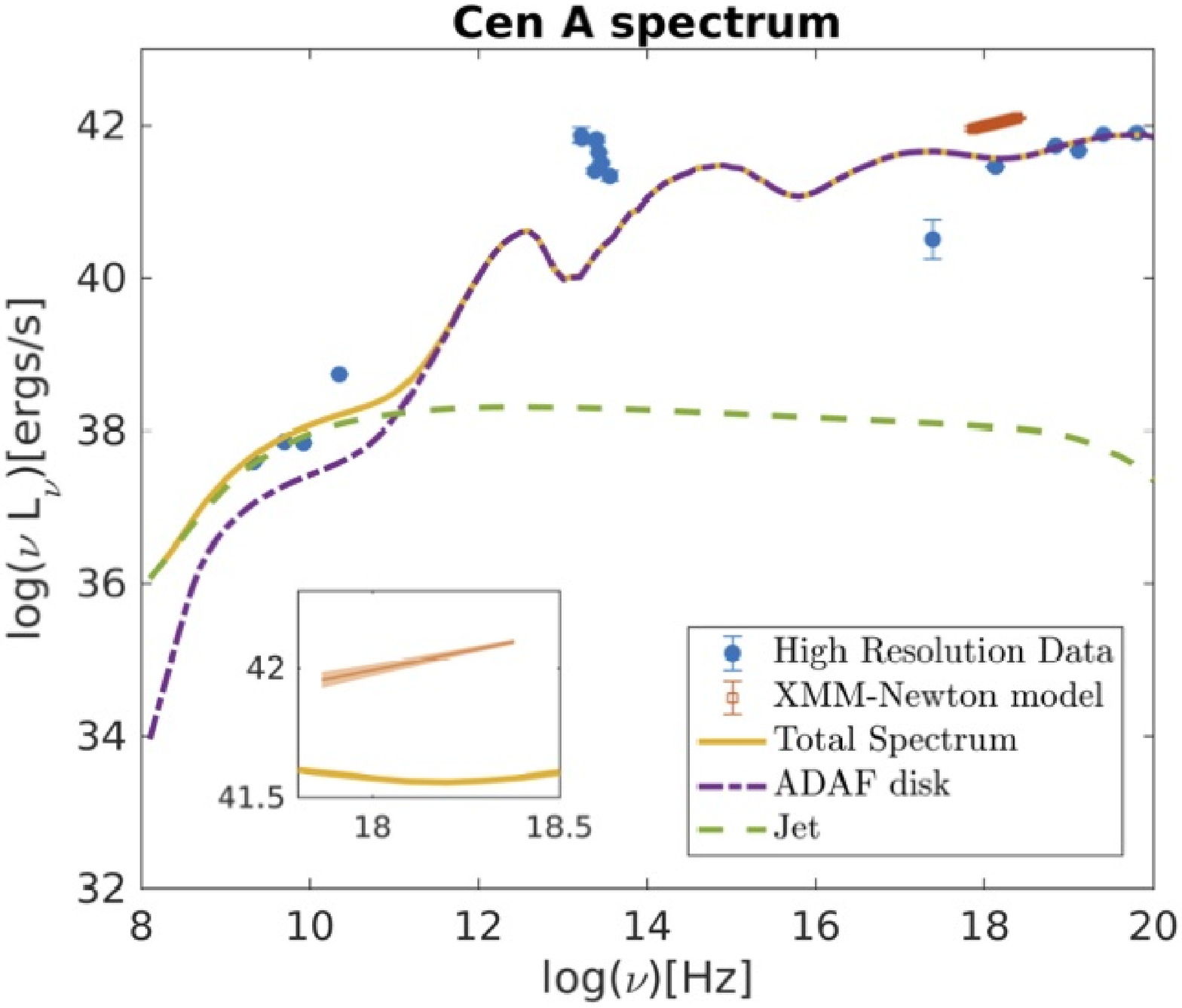}
	\includegraphics[width=2.5in,height=\columnwidth,angle=-90]{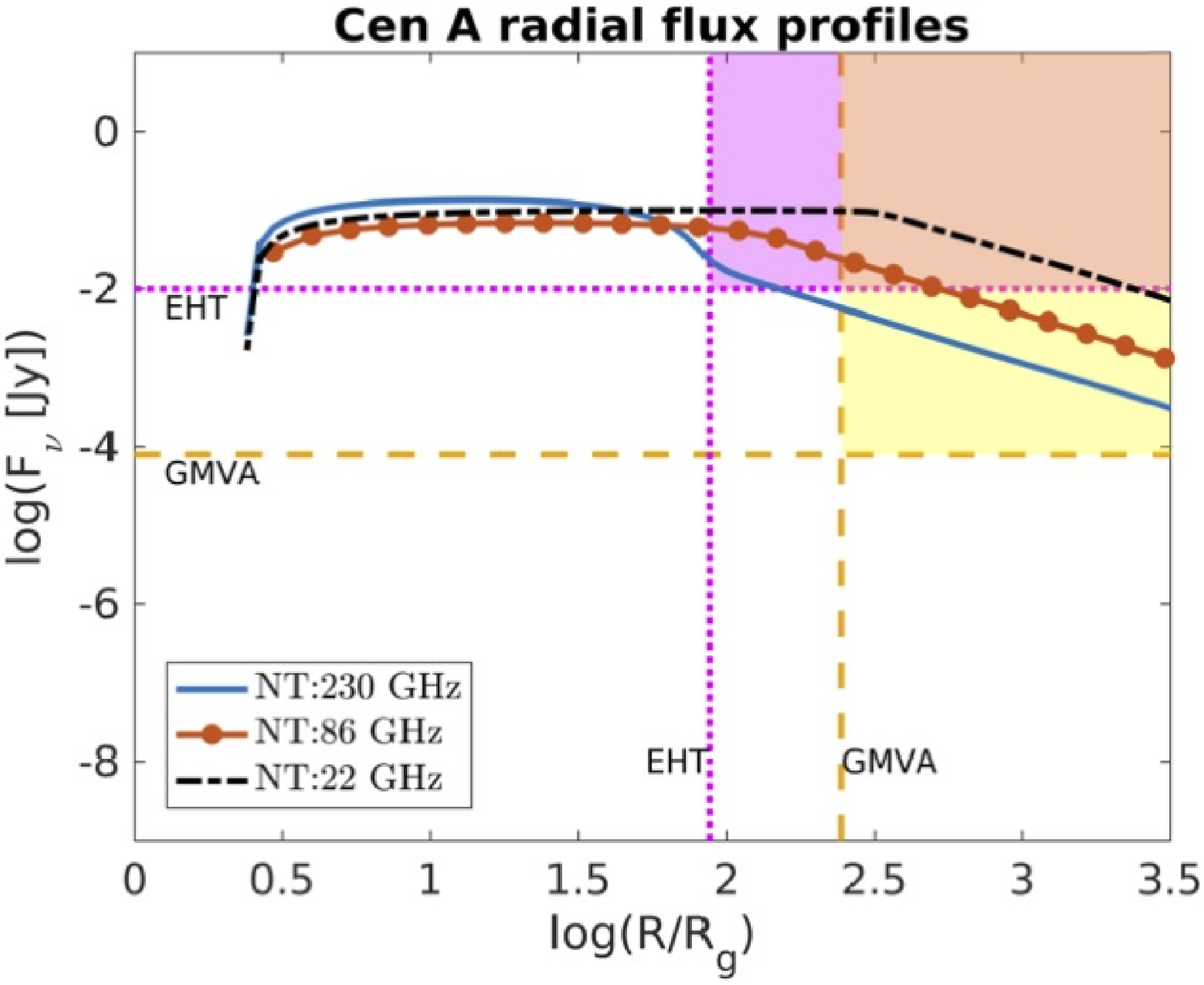}
   \caption{{\it Left Panel}: The multi-wavelength data and model spectrum for Centaurus A. The blue circles corresponds to the high resolution data given in Table-[\ref{tab:CenA}], the red squares are the unabsorbed model X-ray data from XMM-Newton (EPIC-MOS1, August 2013), the purple dot dashed line corresponds to the emission from ADAF with non-thermal electrons, the green dashed line corresponds to the emission from power-law electrons in the jet model and the yellow solid line corresponds to the total emission from both the jet and ADAF models. The inset box displays the XMM-Newton data with the error displayed as the orange shaded region and the model fitted spectra in yellow. {\it Right Panel}: The radial flux profiles with the solid blue curve corresponds to the emission at $230$ GHz, the red circled solid curve corresponds to the emission at $86$ GHz and the black dot dashed curve corresponding to emission at $22$ GHz. The pink and yellow vertical and horizontal lines with the shaded regions are as in the bottom panels of Fig. [\ref{fig:M87}]} 
    \label{fig:CenAspec}
\end{figure*}

\begin{figure*}
	\includegraphics[width=2.5in,height=\columnwidth,angle=-90]{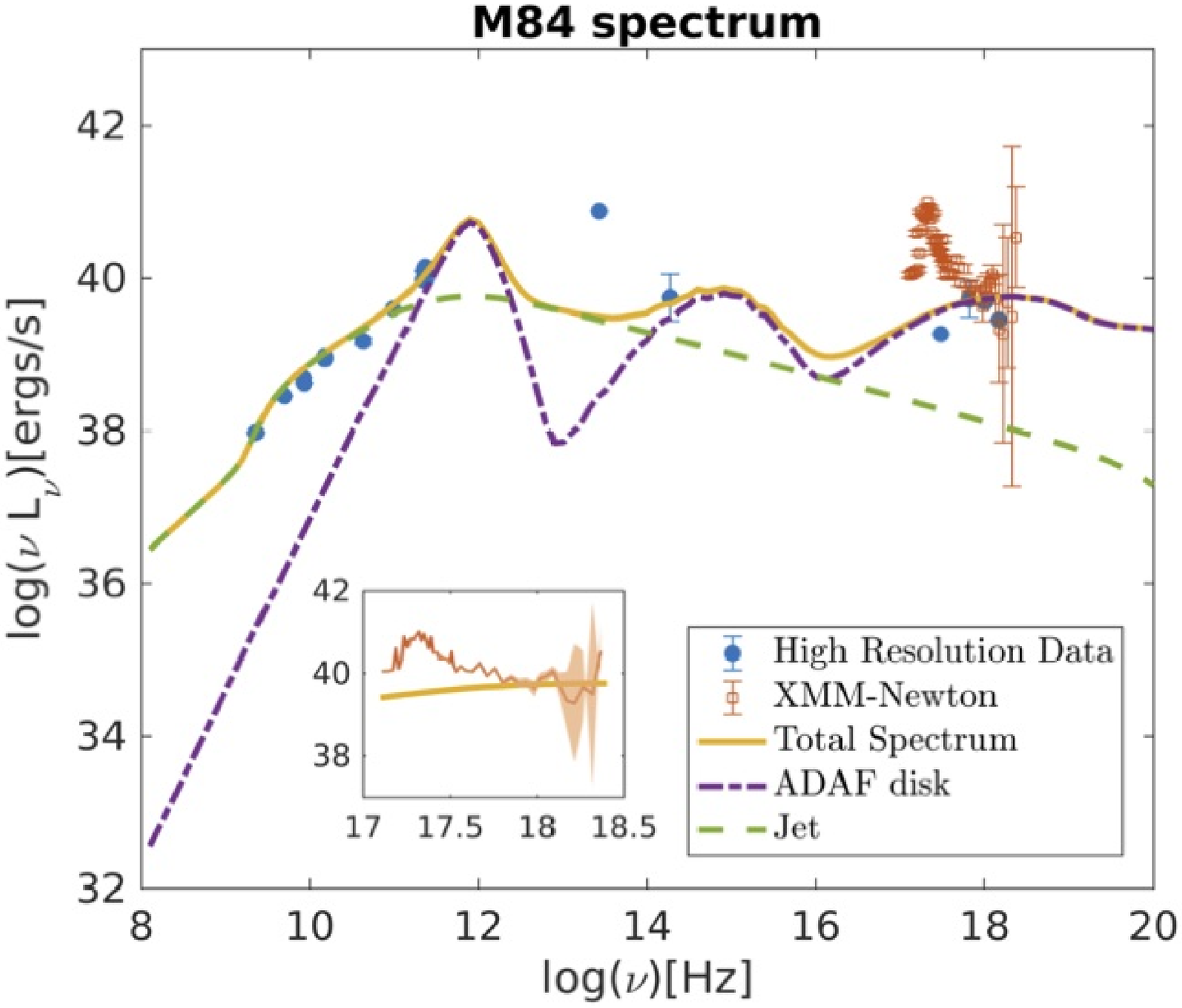}
	\includegraphics[width=2.5in,height=\columnwidth,angle=-90]{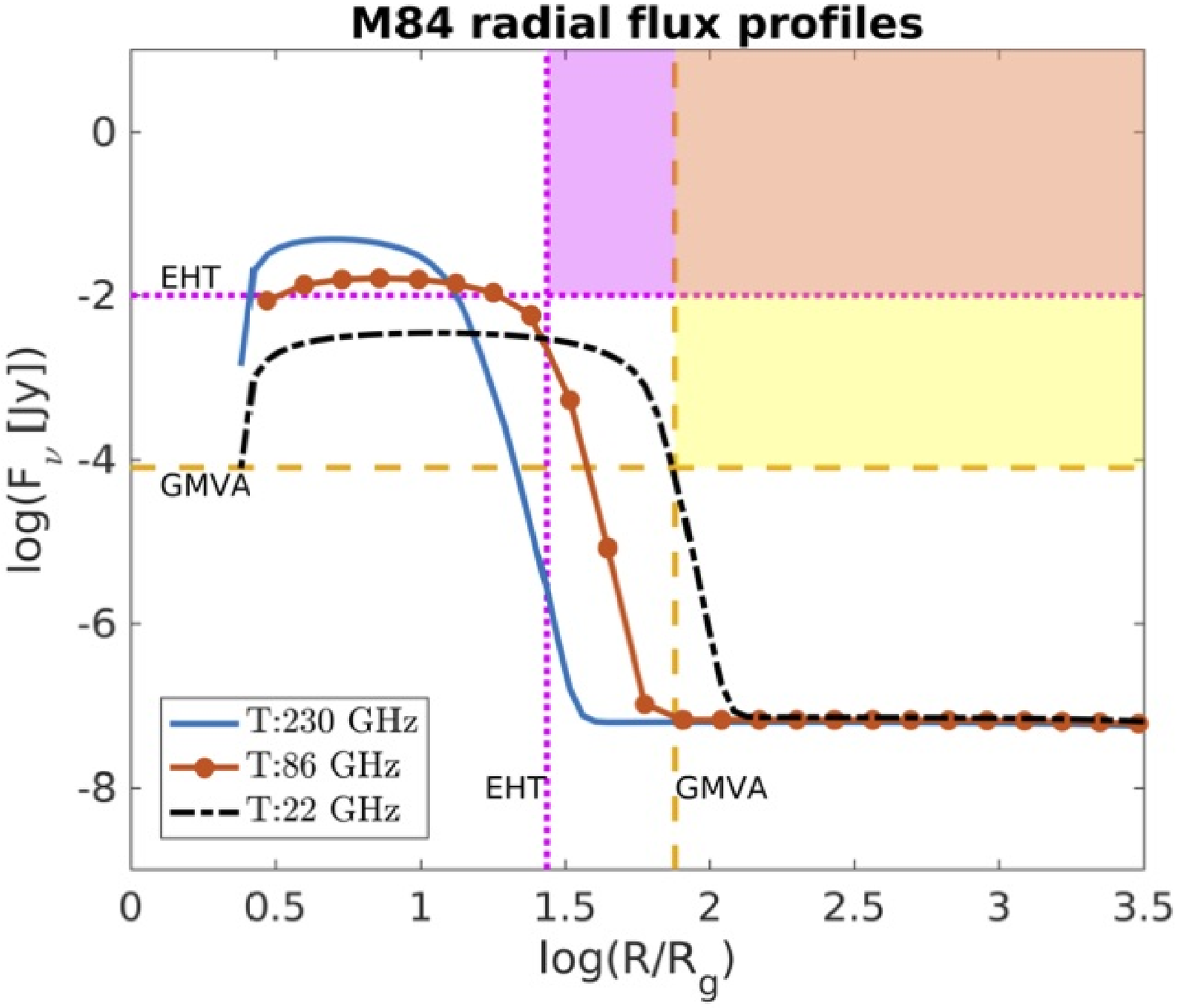}
    \caption{{\it Left Panel}: Same as the left panel of Fig. [\ref{fig:CenAspec}] except that the ADAF emission is only from the thermal electrons. The high resolution data given in Table-[\ref{tab:M84}], the red squares are the X-ray data from XMM-Newton (EPIC-MOS1, June 2011). {\it Right Panel}: Similar to the right panel of Fig.[\ref{fig:CenAspec}]}
    \label{fig:M84spec}
\end{figure*}

\section{Sample and Data}
\label{sec:Sample}
We know that most of the emission from the nuclear region is obtained from the regions in the proximity of the black hole at the center. Thus to model the ADAF accurately and compare the model SED with observation, it is important that we have high resolution (order of sub-arcsec) data from the nuclear regions. It is important to note that in certain bands the flux could be contaminated by star forming regions or an outer thin disk but such regions are much beyond the actual nuclear region. Thus the flux in those bands, such as the IR or optical regions, can be treated as an upper limit while obtaining the model fits. In order to test our model, initially we apply the method to M87 which has been extensively studied and for which now we also have the image of its photon ring with the current EHT observation \citep{Akiyama2019a} which constrains its black hole mass to be $6.5 \pm 0.7 \times 10^9 M_{\sun}$ at a distance of 16.8 Mpc. For this purpose, we have used both the low resolution ($\sim 0.4$ arcsec ) and high resolution ($< 0.2$ arcsec) data obtained for frequencies ranging from radio to X-ray bands from \citet{Prieto2016} (refer to Table 1 and Table 4 of the same work). Their high resolution dataset does not include X-ray data points and hence we have used both datasets from their work.

The primary goal of the EHT Collaboration is to resolve the black hole shadow of, and the bright lensed ring around, supermassive black holes. This goal is currently being pursued with observations of two galactic nuclei - SgrA* and M87. These two targets were selected for having expected black hole shadow and ring size large enough to be resolved by the EHT and for having fluxes greater than 500~mJy at arcsec scales at 230 GHz (a requirement for phasing up ALMA into the EHT; \citet{Matthews2018}). Improvements in the ALMA phasing system is expected to lower the flux limit by factor $\sim$10 in the coming years, thus potentially allowing the resolution of black hole shadows and rings - or at least the accretion flow - in additional nearby galaxies. To this end we have assembled a sample of all (few hundred) nearby galaxies whose black hole mass (either directly measured, or estimated via  the {\it M-$\sigma$} relationship; \citet{Gultekin2009}) implies a ring size larger than 3 microarcsec (Nagar et al., in prep). The typical uncertainties in black hole mass measurements or estimations mean that these all have the potential to be resolved by the EHT. Archival and new high resolution and high frequency observations (Ramakrishnan et al, in prep) are being used to cull this master sample down to the most promising (with respect to milli-arcsec radio cores and relatively high 230 GHz fluxes) candidates for near future EHT observations. In this work, we then focus on five of the most promising EHT targets begond Sgr A* and M87: these are Cen A, M84, NGC 4594, NGC 3998 and NGC 4278. These either have the largest ring diameter (in $\mu$as) or are either radio loud as in the case of Cen A ( Cen A has a smaller ring diameter but has a strong jet which makes it a good candidate to study the accretion flow that can lead to such strong jets). The ring sizes ($10.4 R_{\rm g}$) are estimated from the available distance and mass estimates. The observational parameters which we use for our model are the black hole mass, the distance to the galaxy, the Eddington ratio ($L_{\rm Bol}/L_{\rm Edd}$) and the jet inclination angle are provided in Table -[\ref{tab:Observables}]. The Eddington ratios are obtained either by integrating the SED or using the relation $L_{\rm Bol}=10L_X$ where $L_X$ is the integrated luminosity in the 2-10 keV energy band \citep{Nemmen14}. 

We have obtained high resolution (few arcsec in X-ray to sub-arcsec in radio) nuclear fluxes from the literature for all our sample sources. For most of the galaxies we have considered the data which were measured after the year 2000. The data tables for all these sources have been provided in Appendix-\ref{sec:Data}. The tables list the frequency of observation, the measured fluxes and their errors (in Jy), the resolution of the observations, the instrument and the date of observation (where available) and the references from where the data are obtained. For the sample of five galaxies, we mark these data points using blue circles in the SED plots. It is important to note here that the resolution ($\sim$milli-arcsec) at which the data are taken, is not affected by strong gravitational effects like lensing and hence we can apply our model to fit to these data.  

We have extracted the X-ray spectra of the sources from XMM-Newton, Swift and NuSTAR archival data. From XMM-Newton, we have analyzed observations taken with the EPIC-MOS1 camera (obsID 0673310101 for M84, obsID 0084030101 for NGC4594, obsID 0205010101 for NGC4278, obsID 0790840101 for NGC 3998 and obsID 0724060601 for Cen A) using the Science Analysis Software (\textsc{SAS}, v. 17.0) and the standard analysis threads \footnote{1.https://www.cosmos.esa.int/web/xmm-newton/sas-threads}. We filtered the event lists for high background flaring activity and extracted the source spectrum in circular or annular regions depending on the level of pile-up. The background was extracted from a source-free region in the same CCD. We created the redistribution matrices and the ancillary response files and then grouped the spectra to have at least 20 counts per bin to ensure valid results using $\chi^{2}$ statistical analysis. For NGC 3998 we also analyzed two observations taken with Swift-XRT,  00081893001 and 00081893002. We downloaded the X-ray spectra for each observation from the UK Swift Science Data Centre online tool, which uses the latest version of the Swift software and calibration \citep{Evans2009}. The spectra were grouped to require at least 20 counts per bin using the ftool \textsc{grppha}. For NGC3998 we also downloaded a 103 ks  NuSTAR observation (obsID 60201050002). The data were processed with the tasks nupipeline and nuproducts of the NuSTAR data analysis software version 1.9.3, to extract source and background spectra and construct the corresponding spectral response files.  The same procedure was applied to the data from both detectors, FPM A and B, and their spectra were fit simultaneously. The spectra were analyzed using \textsc{Xspec} version 12.10.0c. We used a simple model consisting of an absorbed power law (\textsc{TBABS*PO}) for all the observations except for M84, for which we used a model for thermal emission from diffuse gas (\textsc{APEC}). This model accounts for the X-ray emission from the intracluster gas that surrounds M84 \citep{Ehlert2013}, so the flux obtained is an upper limit of the X-ray flux from the AGN. To estimate the unabsorbed fluxes for the sources other than M84, we set the column density to zero and use the fluxes from the unabsorbed model. All (absorbed or unabsorbed) X-ray flux data points are marked in the SED plots with red squares. For NGC 3998 we use different symbols to distinguish data from NuSTAR, Swift and XMM-Newton.

\section{Results}
\label{sec:Results}
Our aim in this work is not only to obtain best fit model parameters to the data but also to make a basic prediction (based on these fits) about the type of sources whose accretion flow can be resolved by the EHT or GMVA since these arrays allow observations at high frequencies (high frequency observations not only give a better resolution but in general the ADAF peak is expected to lie in the sub-mm band). Thus in order to make a prediction, we need to obtain the radial flux profile from the ADAF.  We make these predictions for each of our sources at three frequencies 230 GHz, 86 GHz and 22 GHz which are the peak frequencies corresponding to the observing bands of EHT, GMVA and EVN respectively. The radial profile of the emission obtained here gives  the total luminosity for each radial contour  assuming a circular symmetry. %For models considering emission from thermal electrons, the radiation losses are minimum in the outer part of the disk and peaks suddenly at some radius and continues to be flat as we approach the black hole. The flux is maximum at 230 GHz but the peak emission width is smaller while at a lower frequency like 22 GHz the peak emission is much lower but the peak width is wider. In presence of non-thermal electrons, we observe emission through out the disk at these frequencies with a gradual increase towards the center. We have obtained such a profile only for one of the galaxies in our sample {\it i.e.} NGC 4278.

In the following subsection we initially explain the importance of the impact of the model parameters of the ADAF on the total spectrum as well as the radial emission profile. We then further take the example of M87, which has been extensively studied in the literature, to demonstrate the importance of various components in its nucleus to obtain the best fits to the spectrum for two scenarios. Finally we apply the model to the sample of the selected sources and comment whether their ADAF will be resolved by EHT and GMVA using the model, fitted to the available SED data.

\subsection{Effect of Parameters on the Spectrum}
In the following we show how some of the observable parameters and model parameters affect the total spectrum of emission from the ADAF. Fig.[\ref{fig:complots}] and Fig.[\ref{fig:complotsradial}] show the variation in the total spectrum and the emission profile at 86 GHz by varying different model parameters. The top left panel of both the figures, displays the variation of the spectrum by assuming black hole masses ranging from $\sim 10^6$ M$_{\sun}$ (corresponding to a mass similar to that of Sgr A*) to $\sim 10^9$ M$_{\sun}$ (corresponding to the black hole mass of M87). We see that the entire spectrum shifts to higher luminosities for larger masses. In $R_g$ scales, the luminosity is higher for high masses but the radial profile is narrower in these scales than for lower mass ones as can be seen in the radial emission profile. In physical scales, the gravitational radius $R_g$ being larger for black holes with larger masses, the emission profile is wider than the ones for smaller black hole masses. For comparison we have marked the knee of the emission plots for the most massive and the least massive black hole on the top left panel of Fig. [\ref{fig:complotsradial}]. For a black hole of mass $\sim 10^9 M_{\sun}$, the knee corresponds to 0.006 pc whereas for a $\sim10^6 M_{\sun}$ mass, the emission width which appears wider in $R_{\rm g}$ scales, in reality corresponds to $\approx 10^{-5}$ pc in physical scales.   

As discussed, the emission spectrum of the ADAF in general should consist of two peaks in the absence of comptonization. One of them is due to synchrotron emission (peak in the radio band) and the other due to thermal bremsstrahlung emission (peak in the X-ray). It can be seen that the spectrum displays significant differences from the infrared regimes to the X-ray regimes when varying the accretion rate through $\dot{m}_{tr}$ (top right panel of Fig.[\ref{fig:complots}]). This is expected because lower accretion rates lead to lower densities which imply lower electron densities. The thermalization of electrons, which determines the emission in IR and X-ray bands through bremsstrahlung emission and the comptonization of the emitted photons, is not efficient at low accretion rates. For accretion rates corresponding to $\dot{m}_{tr}\sim10^{-5}$, it can be seen that the comptonization of photons is not too efficient. Reduced electron densities also affect the synchrotron emission in the radio regime as shown in a similar plot in Fig.[\ref{fig:complotsradial}]. For higher accretion rates, the luminosity is higher throughout the disk and also the maximum emission region is broader. The outflow parameter $s$ shows similar effects on the spectrum as it also affects the accretion rate (bottom left panel of Fig.[\ref{fig:complots}] and Fig.[\ref{fig:complotsradial}]). Smaller values of $s$ imply lower mass loss rate through outflows. The value of $s$ can vary between 0 and 1; 0 implying a steady accretion flow without any outflow while the value 1 is constrained by energetics \citep{Blandford99, Yuan14}. In the plots here we only consider values of $s$ between 0.1 and 0.3, as all of our galaxies display variations within this range. 

The energy injection parameter $\delta$ on electrons primarily affects the X-ray regime. A higher value of $\delta$ results in a higher luminosity in the X-ray regime which implies higher bremsstrahlung emission through thermal electrons. A similar effect is there in the radio band but the effect is not that pronounced. These effects are shown in the bottom right panel of Fig.[\ref{fig:complots}] and Fig.[\ref{fig:complotsradial}]. The variations in the parameters $s$ and $\delta$ complement each other. A higher value of $s$ results in enhanced outflows which reduces the electron number density, thus reducing the emission. A higher value of $\delta$ implies enhanced emission due to an immense quantity of heat being injected into the electrons.   

From the radial emission plots at 86 GHz, we infer that to resolve the ADAF disk, one needs to target sources with higher mass, higher Eddington ratios and smaller distances.  We further explain this with the example of M87 and then apply it to the sample of our sources.
\begin{figure*}
	\includegraphics[width=2.5in,height=\columnwidth,angle=-90]{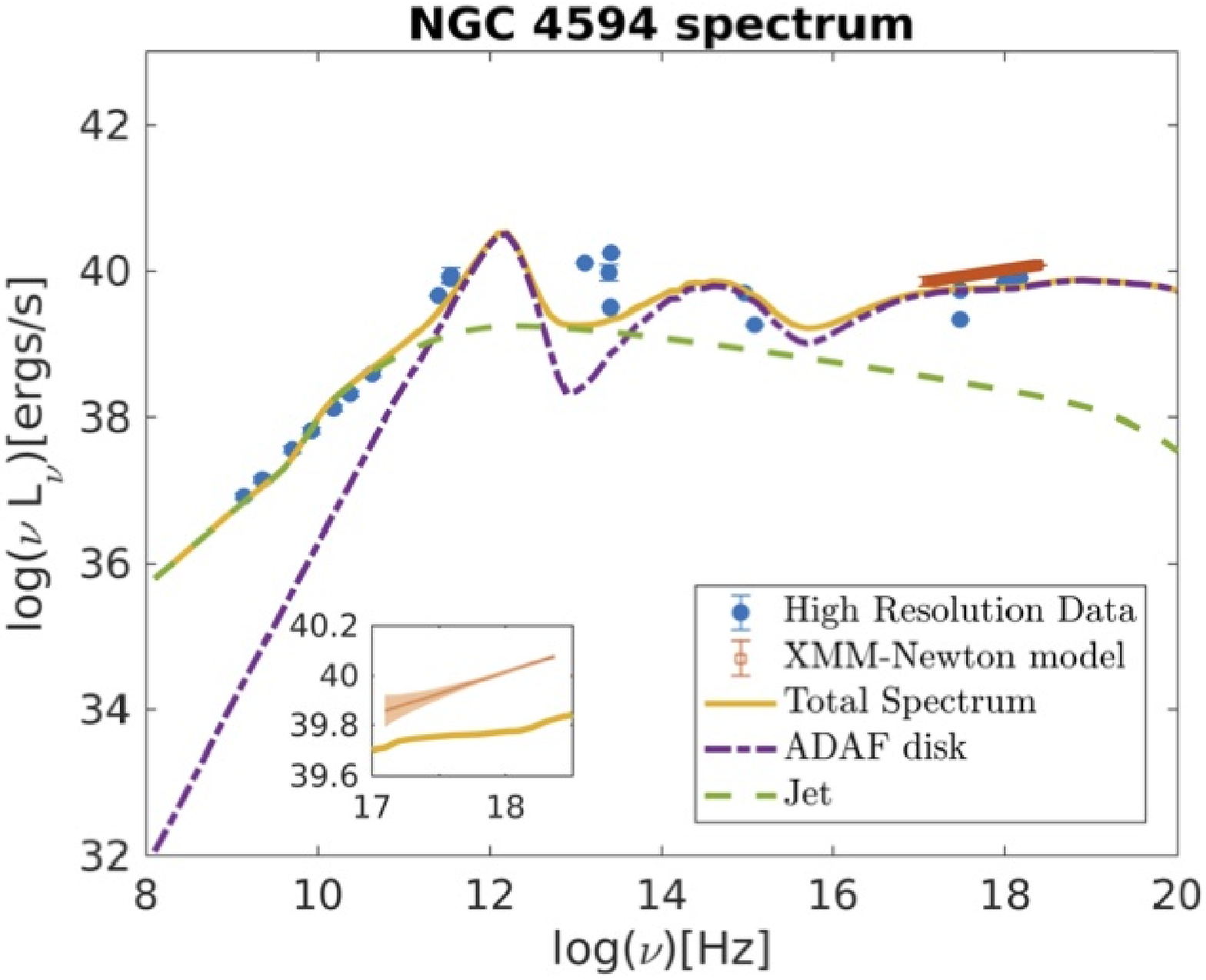}
	\includegraphics[width=2.5in,height=\columnwidth,angle=-90]{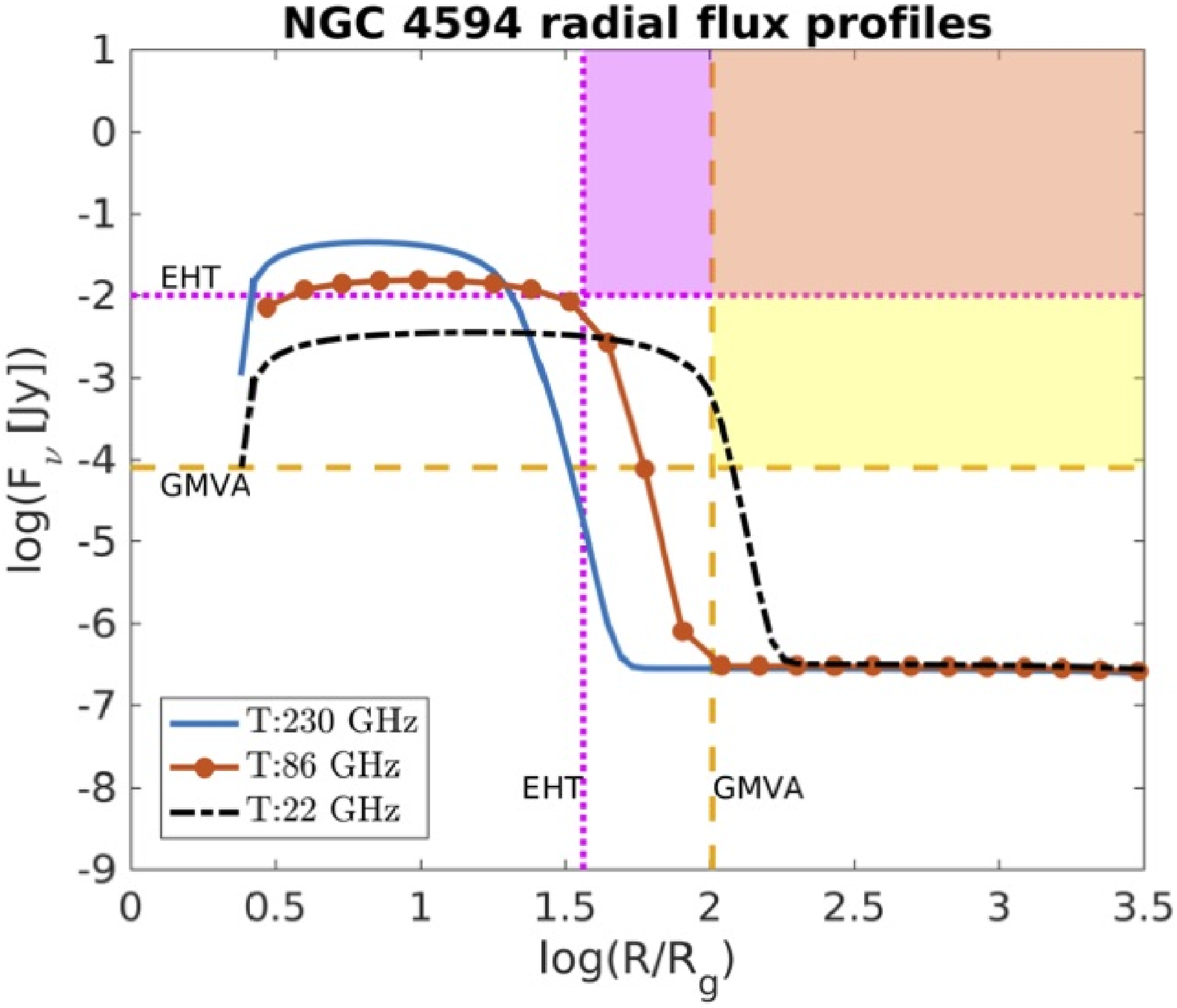}
    \caption{{\it Left Panel}: Same as the left panel of Fig. [\ref{fig:CenAspec}] except that the ADAF emission is only from the thermal electrons. The high resolution data given in Table-[\ref{tab:NGC4594}], the red squares are the unabsorbed model X-ray data from XMM-Newton (EPIC-MOS1, December 2001). {\it Right Panel}: Similar to the right panel of Fig. [\ref{fig:CenAspec}]}
    \label{fig:NGC4594spec}
\end{figure*}

\begin{figure*}
	\includegraphics[width=2.5in,height=\columnwidth,angle=-90]{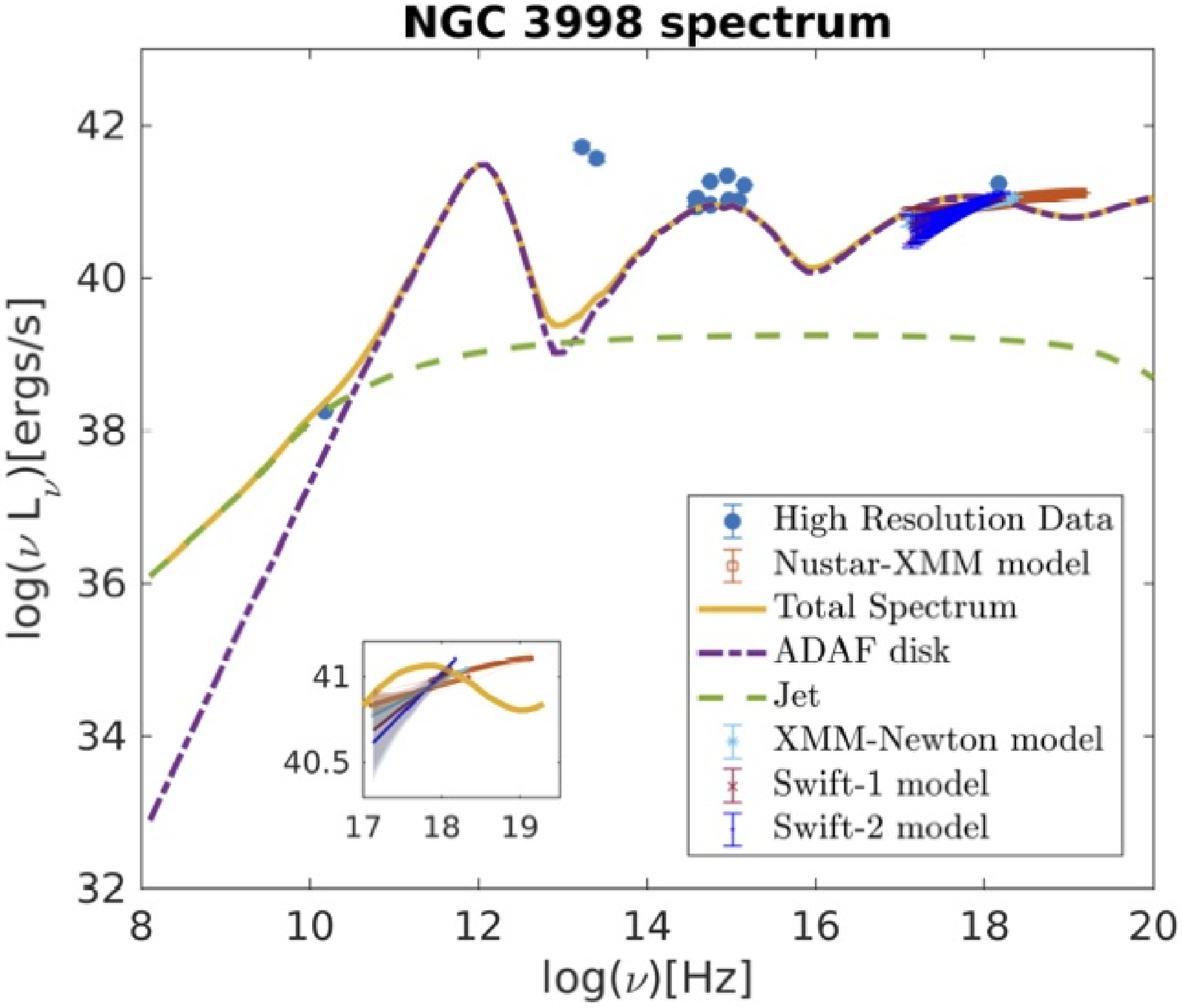}
	\includegraphics[width=2.5in,height=\columnwidth,angle=-90]{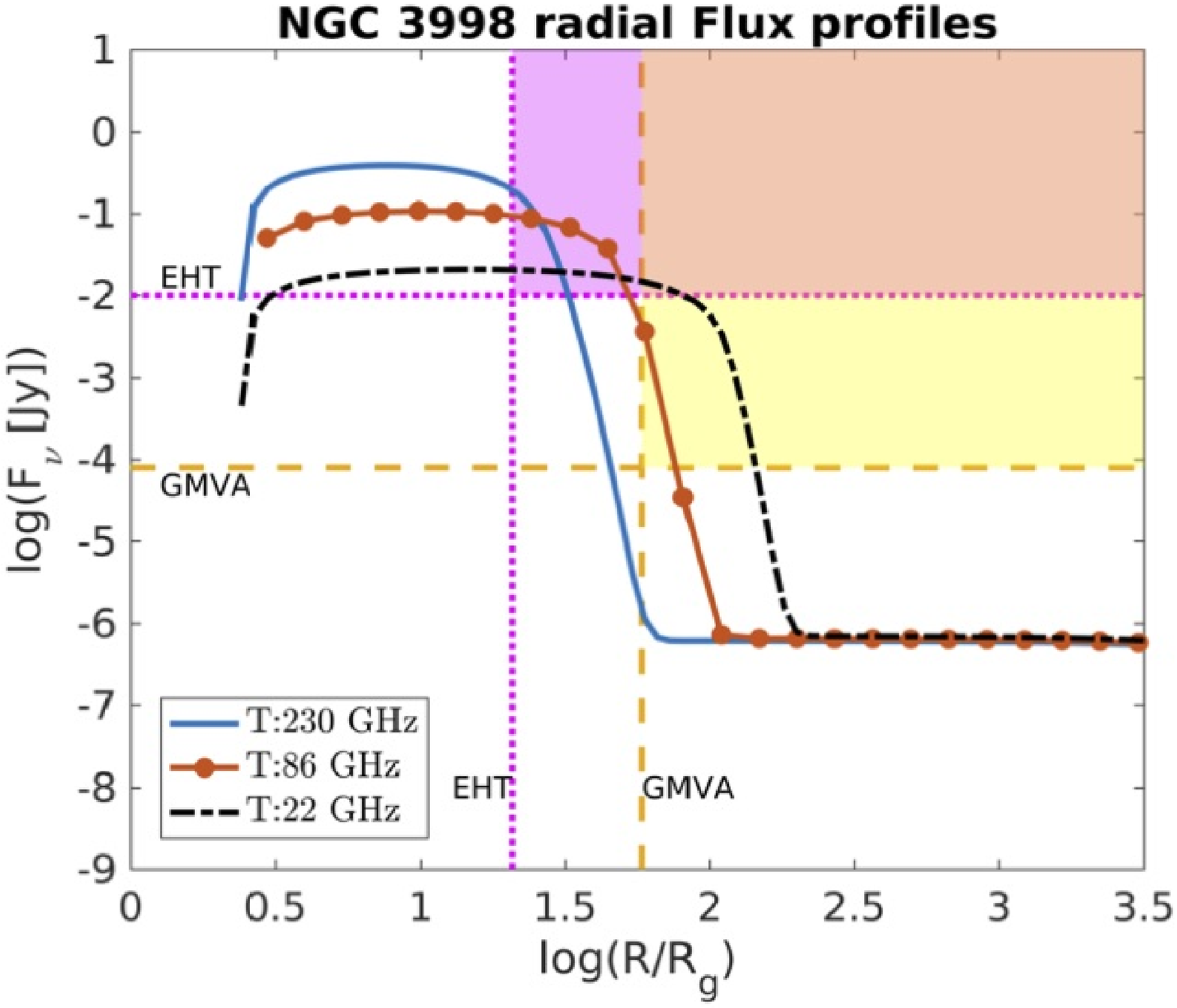}
    \caption{{\it Left Panel}: Same as the left panel of Fig.[\ref{fig:CenAspec}] except that the ADAF emission is only from the thermal electrons. The high resolution data is given in Table-[\ref{tab:NGC3998}], the red squares are the unabsorbed model X-ray data from NuSTAR and XMM (October 25, 2016). In addition we have other X-ray measurements from XMM-Newton (October 26, 2016),  Swift-XRT (October 25, 2016) and Swift-XRT (October 27, 2016) shown with light blue stars, maroon crosses and navy blue dots respectively. {\it Right Panel}: Similar to the right panel of Fig.[\ref{fig:CenAspec}].}
    \label{fig:NGC3998spec}
\end{figure*}

\subsection{Testing the model with M87}
As already mentioned in the section \ref{sec:Sample}, M87 has been extensively studied and modeled considering various scenarios, ranging from SED models to models resulting from GRMHD simulations \citep{Dexter2012, Lu2014, Nemmen14, Akiyama2015, Broderick2015, Russell2015, Monika2017, Li16, Monika2016, Akiyama2017}. It is thus an excellent test case for the model developed here. Although the photon ring has now been resolved by the EHT, the accretion flow is yet to be imaged and hence various estimates on the flow parameters are still uncertain. We have good estimates on the black hole mass and the distance to M87, but we only have indirect estimates on the accretion rate and outflow parameter (which are important factors affecting the SED) given the jet outflow \citep{Broderick2015} and Faraday rotation measurements \citep{Kuo2014}. We consider two scenarios with different accretion rates and outflow parameters which are motivated by the previous studies of \citet{Monika2016, Nemmen14, Li16}. The parameter values, for the models A and B, are given in Table-[\ref{TestM87}]

To compare the resulting SED from our model to observed data, we use both the high and low resolution data from \citet{Prieto2016}, which covers the complete spectral band. The mass and distance estimates are obtained from the results of \citet{Akiyama2019a}. We take into account three major components which contribute to the SED significantly: The ADAF consisting of only thermal electrons, ADAF including also the non-thermal electrons, and the jet. We show these three components in the top panels of Fig.[\ref{fig:M87}] with the black dotted line, purple dotted dashed line and the green dashed line, respectively. In addition to synchrotron emission by the thermal electrons, the synchrotron emission from the non-thermal electrons results in an enhancement in the luminosities at low frequencies, but the effect of self absorption in the ADAF still exists. The comptonization of photons by these electrons then enhances the flux from IR to X-rays. The best fit to the radio data is obtained when we also consider the synchrotron emission from the jet. For M87, the best fit model to the entire data set is when we consider an ADAF consisting of thermal plus non-thermal electrons and a jet. The inner jet of M87 has been constrained to an inclination angle of 10 degrees and $\Gamma_j=6.0$ \citep{Nemmen14}. We deduce that Model A almost perfectly matches the data but the fitted accretion inflow rate is $\dot{m}_{42R_g}=2.4 \times 10^{-4}$ (Eddington ratio at 42 $R_g$) which is an order of magnitude larger than the value obtained by \citet{Kuo2014} ($\dot{m}_{42R_g}=7 \times 10^{-6}$) with their Faraday rotation measurements. Our fitted value however is similar to that deduced by \citet{Monika2016} with GRMHD simulations. As can be seen in the top left panel of Fig.[\ref{fig:M87}], the X-ray data points fit the emission from the ADAF. Model B on the other hand shows a larger deviation from the measured data points and the X-ray data points better fit the emission from the jet but we obtain $\dot{m}_{42R_g}=2.3 \times 10^{-5}$, which is a good approximation to observed value of \citet{Kuo2014}. The model B parameter values of $\dot{m}_{tr}$, $\delta$ and the outflow parameter $s$ are obtained from \citet{Li16}, where they use a lower black hole mass and hence the values of other parameters are adjusted to obtain a better fit to the data. Faraday rotation measurements depend on the strength of magnetic field which is quite model dependent and hence also the resulting value of $\dot{m}_{42R_g}$. 

The radial flux profile of the ADAF at 230 GHz, 86 GHz and 22 GHz, obtained with these parameter values, are shown in the bottom panel of Fig.[\ref{fig:M87}] for both models A and B. We show the cases with and without considering the emission from non-thermal electrons. The presence of non-thermal electrons causes a gradual rise in the flux (with decreasing radius) for the entire flow whereas their absence causes a steep rise in the flux only in the innermost regions of the flow. Both the figures in the bottom panel clearly show that the accretion flow in M87 should be well resolved with the EHT for both the models A and B. It is to be noted that the emitted flux is higher and broader in case of model B even though the densities are lower (smaller accretion rate and greater outflow compared to model A) because the energy injected into the electrons is higher (higher value of $\delta$). This leads to greater emission. The dynamic range and the fidelity of the published image \citep{Akiyama2019a} is not good enough to trace this flow or ensure which of the models is better but near future observations, with better $uv$ spacing, higher bandwidth and more sensitivity will likely do so.

\subsection{Model fits to the sources.}
After testing the robustness of our model with the multi-wavelength data of M87, and demonstrating the difference and importance of the emission from the various components which contribute to the total nuclear flux at different energy bands, we then use this to our sample of five galaxies and obtain the parameter values which fit the data. We note that since these objects do not emit efficiently, their accretion rates cannot be estimated directly from their luminosities. Thus other methods, like Faraday rotation measurements must be used to estimate the mass accretion rates. Our modeling predicts a luminosity given the accretion rate and other model parameters, which can be compared to the observed luminosities a posteriori. Since for our sample of galaxies, we do not have accurate Faraday rotation measurements or other direct constraints on the accretion rate in the inner accretion flow, we allow the accretion rate and other parameters to freely vary in order to obtain the best fit. The only fixed quantity is the black hole mass and distance (Table-[\ref{tab:Observables}]). To judge the goodness of fit to the data we use eye estimates (as also in \citet{Nemmen14, Li16}, etc.) and not any automatic fitting techniques, like maximum likelihood, method of least square or a chi-squared test, because of the complications involved in obtaining solutions to the dynamical equations of the ADAF and the radiative transfer code. In addition there are uncertainties pertaining to the coupled effects of various parameters. Using automated processes would be computationally expensive and is not necessary given the other uncertainties involved. Unlike M87, most of these sources have not been studied so extensively. There are estimates on the accretion rate through X-ray observations \citep{Russell2013, Rafferty2006} where an estimate of the Bondi accretion is made. Our accretion rate estimates agree with \citet{Nemmen14}. They have shown that, within the error limits, these estimates agree with the above observations. The model parameters used to obtain the SED which fit the high resolution data for each of the sources are given in Table [\ref{tab:Observables}] and [\ref{tab:Parameter}]. With these parameter values, the radial flux profiles for the ADAF of each of these sources are obtained and tabulated in Table [\ref{tab:Res}]. These are then used to make a prediction about detecting and resolving the accretion flow of these sources.  Following we discuss the results for each of these sources.
 
\subsubsection{Cen A/ NGC 5128}
NGC 5128, popularly known as Centaurus A or Cen A is one of the nearest galaxies at a distance of $3.8$ Mpc \citep{Harris2010}. We have used a black hole mass $M_{\rm BH}\sim 5.5 \times10^7 M_{\sun}$ \citep{Cappellari2009}. The expected ring diameter ($\sim 10.4 R_g$) using the mass and distance estimate is $\theta_{\rm Ring}=1.5 \mu$as. Although the ring size of Cen A is quite small, it is one of the sources targeted by the EHT as it is radio loud due to its strong jet. Hence it would be of interest to make a detailed observation to understand the kind of accretion flow which leads to such strong jets. The observed and modeled SED are on the left panel of Fig.[\ref{fig:CenAspec}]. The unabsorbed spectral data points (red squares) correspond to the observation by XMM-Newton in August 2013. The radio observations have a higher resolution but are older (Table- \ref{tab:CenA}). We concentrate on fitting the model to the data at low as well as high frequencies. To fit the data in the high frequency bands, it is also important to consider the emission from the non-thermal electrons in the ADAF as was also the case for M87. With the current fits, the peak emission region is resolvable but not observable by the EHT or the GMVA as can be seen in the right panel of Fig.[\ref{fig:CenAspec}] but the emission from the non-thermal electrons from the outer parts of the flow may be resolved and detected by the EHT and GMVA in the 8 hour integration time. Obtaining a recent high resolution data set in the radio band will allow us to make a better estimation of the SED. 

\begin{figure*}
	\includegraphics[width=2.5in,height=\columnwidth,angle=-90]{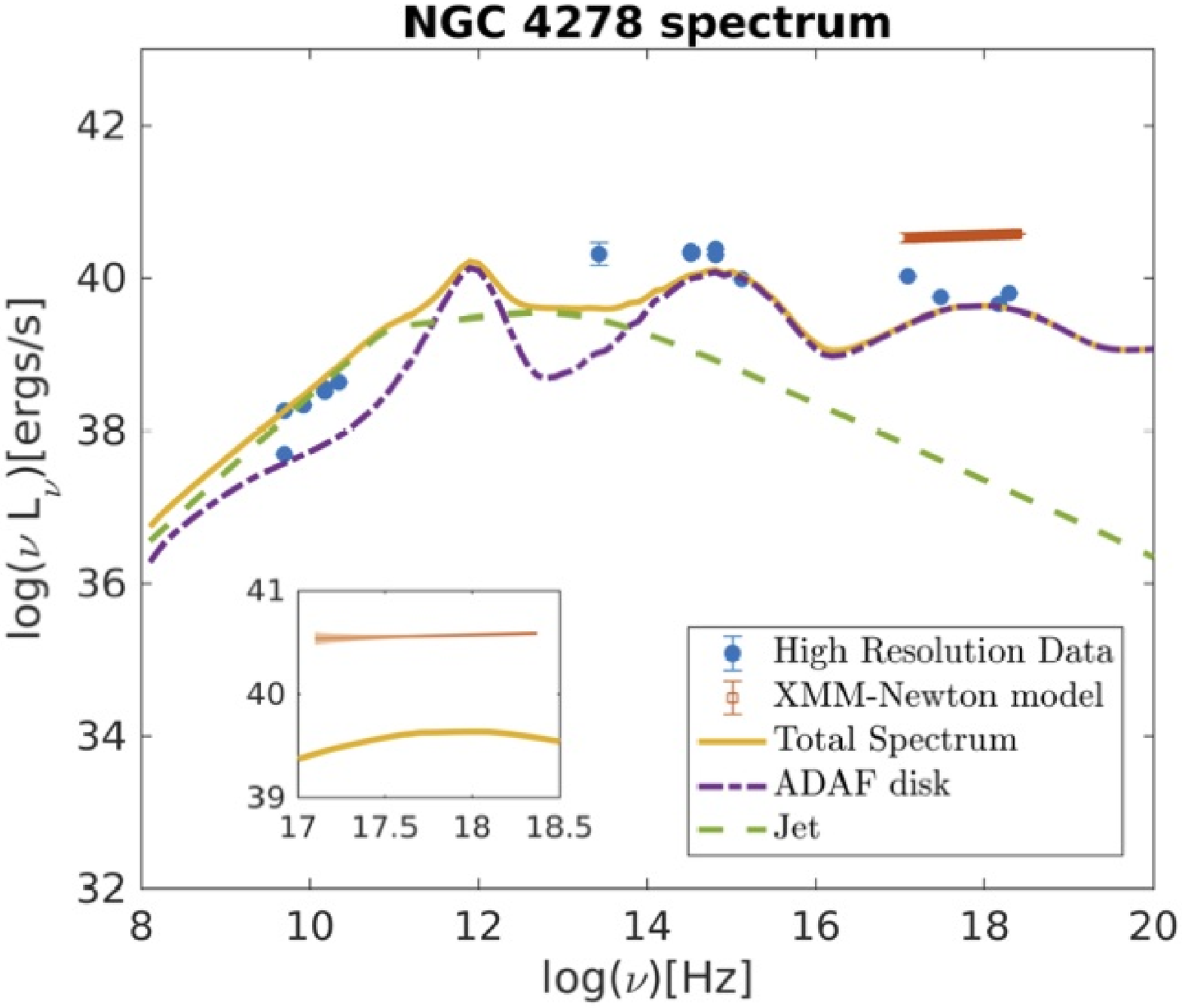}
	\includegraphics[width=2.5in,height=\columnwidth,angle=-90]{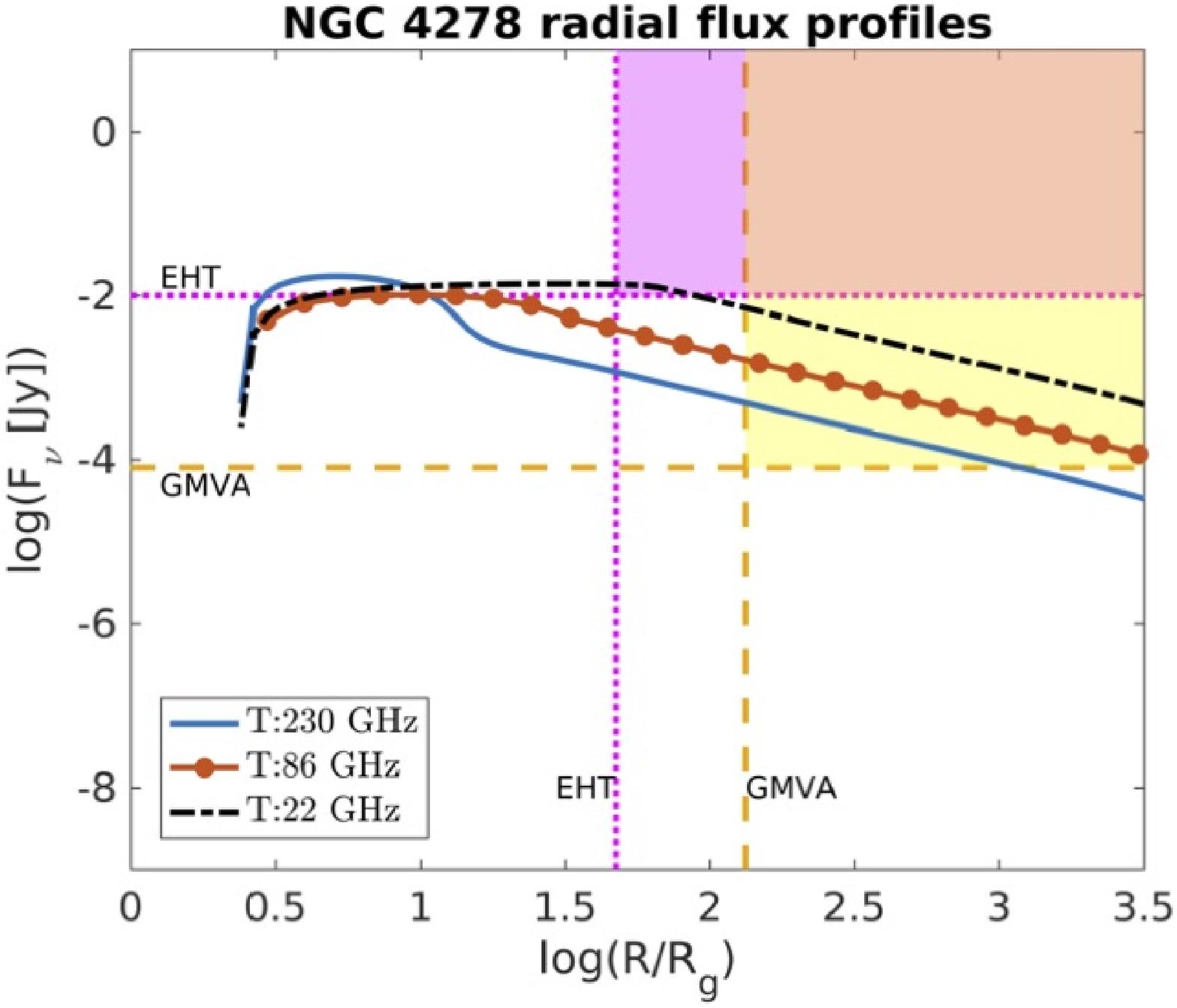}
    \caption{{\it Left Panel}: Same as the left panel of Fig. [\ref{fig:CenAspec}]. The high resolution data (blue circles) is given in Table-[\ref{tab:NGC4278}] while the red squares are the unabsorbed model X-ray data from XMM-Newton (EPIC-MOS1, August 2013). {\it Right Panel}: Similar to the right panel of Fig.[\ref{fig:CenAspec}]}
    \label{fig:NGC4278spec}
\end{figure*}

\subsubsection{M84/ NGC 4374}
NGC 4374 also know as Messier 84 or M84 belongs to the Virgo cluster. It is at a distance of $17.1$ Mpc with a central black hole of mass $=7.94 \times 10^{8} M_{\sun}$. The estimate on the central black hole of M84 is quite uncertain \citep{Walsh2010, Bower1998} and hence the ring size quoted here is expected to improve with better estimates on its mass. It displays small variations in flux at 220 GHz and 230 GHz over a period of a few days as reported by \citet{Bower2017}. The nuclear flux with high resolution observation is tabulated in Table - [\ref{tab:M84}]. The XMM-Newton data were observed on June, 2011 and display an emission like feature. As mentioned by \citet{Donato2004}, such features in X-ray could be due to the presence of a point source in the vicinity of the central black hole. As already mentioned this feature could also be due to the thermal emission from the intracluster gas surrounding M84 \citep{Ehlert2013}. The data so obtained favor an ADAF model spectrum with emission only from thermal electrons. From the radial profile of emission flux, the M84 ADAF appears to be partially resolvable by the EHT but not observable either with the EHT or with the GMVA as seen in the right panel of Fig.[\ref{fig:M84spec}].

\subsubsection{Sombrero/ M 104/ NGC 4594}
NGC 4594, also popularly called the Sombrero galaxy or M104 is a spiral with a LINER nucleus hosting a black hole with mass $M_{\rm BH}=3.2 \times 10^{8}$. It is at a distance of $9.1$ Mpc \citep{Mezcua2014, Prieto2010, Nemmen14}. With these estimates on mass and distance, we obtain the ring size $\theta_{\rm Ring}\sim 3.6 \mu$as. The galaxy is almost edge-on and is crossed mid-plane by a distinctive dust lane. It was undetected at 11.8 $\mu$m and thus the mid-IR emission probed has  to  be  largely  diffuse,  escaping  detection  in  ground-based observations. The AGN contribution, if any, is thus tiny at these wavelengths \citep{Reunanen2010}. The left panel of Fig.[\ref{fig:NGC4594spec}] shows the SED with the observed data points as tabulated in Table -[\ref{tab:NGC4594}]. The XMM-Newton data were observed on December 2001. The source appears stable over a larger period of time at all bands. We obtain a reasonable fit to the data points with an ADAF with emission only by thermal electrons. Similar to M84, the Sombrero ADAF disk is partially resolvable by the EHT but not observable either with the EHT nor with GMVA as seen in the right panel of Fig.[\ref{fig:NGC4594spec}].

\subsubsection{NGC 3998}
NGC 3998 is a nearby (Distance$=13.1$ Mpc) S0 galaxy located at the outskirts of the Ursa Major group. The galaxy hosts a low power radio AGN with a flat spectrum radio core. The central black hole has a mass $M_{\rm BH}=7.9 \times 10^8 M_{\sun}$. This galaxy has been studied well with various probes in the X-ray using Swift, XMM-Newton and NuSTAR shown with various markers in the left panel of Fig.[\ref{fig:NGC3998spec}]. These observations were taken between October 25 to October 27, 2016. The other high resolution data points from other frequency bands in the figure is tabulated in Table -[\ref{tab:NGC3998}]. The galaxy has not been observed in the radio bands very extensively in the recent past and hence only one data point in the radio is considered. With the given data points, we obtain a good fit to the data when the emission is from thermal electrons in the ADAF. The model of the jet emission here predicts a weaker contribution to the emission from the jet than from the ADAF. With this model, the ADAF appears to be observable both by the EHT but partially by the GMVA as the right panel of Fig.[\ref{fig:NGC3998spec}] shows.

\subsubsection{NGC 4278}
NGC 4278 is a LINER source displaying variability both in short and long timescales in X-ray and radio bands \citep{Younes2010}. It displays a two-sided structure, with symmetric S-shaped jets emerging from a flat-spectral core \citep{Giroletti2005}. Its nucleus consists of a black hole of mass $M_{\rm BH}\sim 4 \times 10^8 M_{\sun}$ and the galaxy is at a distance of 14.9 Mpc \citep{Giroletti2005, Nemmen14}. The data in Table -[\ref{tab:NGC4278}] displays variability which can also be seen in the left panel of Fig.[\ref{fig:NGC4278spec}]. The XMM-Newton data points are taken from the observation of August 2013. The jet inclination angle is between $2 < i < 4$ degrees. With all the uncertainties taken into account, we fit the data with an ADAF which consists of emission also by the non-thermal electrons as we have seen for M87 and Cen A. It can be seen in the right panel of Fig.[\ref{fig:NGC4278spec}] that the presence of power-law electrons results in emission through the entire ADAF at the radio frequencies although displaying a gradual increase towards the core. Given such an emission profile, it may be difficult to observe the source with the EHT due to the low flux at 230 GHz but could possibly be observed by the GMVA.

\begin{table*}
% 	\centering
	\caption{The table displays if the sources are detectable and resolvable by the EHT or the GMVA in an 8 hour integration time.}
	\label{tab:Res}
	\begin{tabular}{lcccr} % four columns, alignment for each
		\hline
		Source &Resolvable (EHT) &Detectable (EHT)&Resolvable (GMVA)&Detectable (GMVA) \\
		\hline
		Cen A & Yes &Only a part of the outer region &Outer regions& Only the outer regions\\
		M84 &Partially & No & No &No\\
 		NGC 4594 & Partially & No &No &No\\
 		NGC 3998 & Yes &Yes &Partially &A part of the outer region\\
 		NGC 4278 &Outer regions &No &Outer regions &Outer regions\\
		\hline
	\end{tabular}
\end{table*}

\section{Summary and Discussion}
\label{sec:Discussion}
Our primary aim in this work was to device a simple method to predict whether the accretion flow in the nucleus of some of the nearby galaxies will be observable and resolvable using VLBI techniques. Also \citet{Nemmen14} performed a similar SED fitting to some of the galaxies in our sample but here in addition we also compare the scenario where the ADAF has emission due to the presence of non-thermal electrons to the one without non-thermal electrons. It can be seen that for some of the sources in our sample show a better fit to the data in the presence of non-thermal electrons. With this work we not only intend to obtain the model parameter values which fit the data but also proceed a step forward in using those parameter values to obtain the radial flux profile which is then compared with the EHT and GMVA resolution and RMS sensitivities, thus enabling us to predict whether these telescopes can resolve the ADAF. To compare the SED, we obtained the latest data with highest resolution available from the literature. Since the data range from milli-arcsec scales to a few arc second scale, we have included in our model, all the components (ADAF with and without non-thermal electrons and a jet component) that could affect the flux at these resolutions.

The important conclusions that we arrive at through this investigation are as follows:
\begin{itemize}
    \item To understand the importance of various parameters in the ADAF model which impact the SED, we performed a methodical study by varying these parameters and observed the variation in SED as well as the radial emission (only from thermal electrons) profiles as shown in Fig. [\ref{fig:complots}] and Fig. [\ref{fig:complotsradial}]. We thus infer from this investigation that while targeting sources for observation, it is important to target those sources with high black hole masses and higher Eddington ratios. This ensures a wider profile of emission in physical scales.
    
    \item We tested this framework with the SED of M87 considering models A and B. The basic motivation to select these models was to consider different accretion rates and outflow parameters which compare to the values in literature \citep{Nemmen14, Li16, Monika2016}. With these models, we obtained the best fit to the data with a model consisting of the Jet and an ADAF with thermal plus non-thermal electrons. Model A fits the data better and the accretion rate is similar to the result of a GRMHD simulation by \citet{Monika2016} but the accretion rate obtained using model B is in accordance with the Faraday rotation measure accretion value \citep{Kuo2014}. Since both the models provide radial profiles which are within the observable regime of the EHT, future EHT observations may help to distinguish the two scenarios. 
    
    \item We then use this model to obtain the model parameters for each of the 5 sources in our sample of galaxies by comparing the modeled SED with the observed data. Two of these sources (Cen A and NGC 4278) fit the data better when the emission from non-thermal electrons is also considered in the ADAF. These sources thus demonstrate a smooth radial profile in contrast to the others (M84, NGC 4594 and NGC 3998) which favour an ADAF consisting of only thermal electrons. Although we may not be able to resolve the region of maximum emission from the ADAF for Cen A but the flow can still be partially observed at higher radii due to the flux from non-thermal electrons. Table [\ref{tab:Res}] summarizes these predictions.
    
    \item With our model fits, we find that the radial profile of NGC 3998 is expected to be resolved very well with both EHT and GMVA. The ADAFs of M84 and NGC 4594 may be fairly resolved by EHT and not with GMVA, but may not be observable within the current flux limit of the EHT. To observe these, we need better sensitivities of the telescope and longer integration times.
\end{itemize}

We would like to emphasize here the fact that the prediction about resolving the ADAF using VLBI techniques is model dependent and may vary with variation in the estimation of the black hole mass and distances. Also strong gravitational lensing may lead to a variation in the width of the emission profile but such effects are beyond the scope of this work. For sources which exhibit variability, it is important to observe those sources to obtain the SED over a period of time which is shorter than their variability period. This would then allow to constrain the model parameters better.

The simple model considered here consists of various parameters, some of which can be degenerate. Hence making accurate predictions about the various model parameters is difficult.  A few of the model parameters that we have fixed in this kind of 1-D model will naturally evolve in a 2-D or 3-D MHD or GRMHD simulations. Such simulations would thus be able to produce a more realistic picture with turbulence and other additional features but such simulations are computationally expensive and simulating such a model for each source for an initial study for model predictions can be time-consuming. Thus the simple method of modeling the ADAF that we have used here can be used as a first approach to check if realistic models fit to the data and to provide a preliminary sense of their predictability. Having a first idea and targeting a good source, one can then resort to 2-D and 3-D simulation of such systems as has been pursued in case of M87 and Sgr A* \citep{Davelaar2018, Monika2016, Monika2009}. A higher order simulation may naturally lead to the existence of non-thermal electrons in the disk through turbulent flows, magnetic re-connections and other processes.

The tangible evidence of the parsec-scale structure and the brightness temperature constraints of these sources that complement the results obtained in this work is currently being studied using VLBI observations. Note that, for our detection limits in figures displaying the radial plots for our sources and Table-[\ref{tab:Res}], we use the expected rms in a map resulting from an 8 hour observation with the EHT or GMVA. Our true detectability threshold is likely to be even better (i.e. lower) as the  the model radial profiles shown are the summed fluxes in radial annuli. However this would depend on many factors including image fidelity, $uv$ coverage, and inclination of the accretion flow to our line of sight.

\section*{Acknowledgements}
We acknowledge the following funding sources: CONICYT Programa de Astronom\'ia Fondo ALMA-Conicyt 2016 31160001, the Conicyt PIA ACT172033 as well as the BASAL Centro de Astrof\'isica y Tecnolog\'ias Afines (CATA) AFB-170002. BB thanks funding via Fondecyt Postdoctorado (project code 3190366). FGX is supported in part by the National Program on Key R\&D Project of China (Grants 2016YFA0400804), the NSFC (grant 11873074), and the YIPA program of CAS (id. 2016243). NMN also acknowledges support from Conicyt through Fondecyt 1171506.

%%%%%%%%%%%%%%%%%%%%%%%%%%%%%%%%%%%%%%%%%%%%%%%%%%



%\bibliographystyle{mnras}
%\input{BB_aph}
\bibliographystyle{mnras}
\bibliography{BB_aph}

%\begin{thebibliography}{99}
%\end{thebibliography}

%%%%%%%%%%%%%%%%%%%%%%%%%%%%%%%%%%%%%%%%%%%%%%%%%%

%%%%%%%%%%%%%%%%% APPENDICES %%%%%%%%%%%%%%%%%%%%%

\appendix
\label{Appendix}
\section{Data Tables}
\label{sec:Data}
We tabulate the high resolution data collected from literature for our sample of galaxies NGC 5128, NGC 4374, NGC 4594, NGC 3998 and NGC 4278. These data points are marked in blue circles in the multi-wavelength spectrum of these sources in the main text. The X-ray data for these samples are from XMM-Newton. For NGC 3998, the X-ray data is also available from the observation of NuStar as well as Swift detectors. The details about the X-ray data is already mentioned in section [\ref{sec:Sample}] and thus not explicitly mentioned here.
\begin{table*}
	\centering
	\caption{High resolution data for Cen A/NGC 5128. The columns here correspond to frequency in Hz, flux in Jansky, error-bars in Jansky, resolution of the instrument in arcsec, Instrument of observation/ time of observation if available and the references where the data are reported.}
	\label{tab:CenA}
	\begin{tabular}{lccccr} % four columns, alignment for each
		\hline
		$\nu$ (Hz) & Flux (Jy) & Error (Jy) & Resolution (arcsec) & Instrument & Reference\\
		\hline
		$2.2 \times 10^9$ & $1.03$ & - & $10^{-3}$ & VLBA(1999) & \citet{Tingay1998}\\
		$5.0 \times 10^9$ & $8.3 \times 10^{-1}$ & - & $10^{-3}$ & VLBA(1999) & \citet{Tingay1998}\\
		$8.4 \times 10^9$ & $4.8 \times 10^{-1}$ & $4.5 \times 10^{-4}$ & $0.5 \times 10^{-3}$ & VLBI & \citet{Muller2011}\\
		$2.23 \times 10^{10}$ & $1.4$ & $1.2 \times 10^{-3}$ & $1.3 \times 10^{-3}$ & VLBI & \citet{Muller2011}\\
%		$2.30 \times 10^{11}$ & $5.5 \times 10^{-1}$ & $1.0 \times 10^{-1}$ & - & EHT & Ask Neil\\
		$1.64 \times 10^{13}$ & $2.6$ & $6.5 \times 10^{-1}$ & $0.53\times 10^{-3}$ & - & \citet{Radomski2008}\\
		$1.67 \times 10^{13}$ & $2.3$ & $2.9 \times 10^{-1}$ & $<0.4$ & - & \citet{Asmus2014}\\
		$2.38 \times 10^{13}$ & $6.0 \times 10^{-1}$ & $5.0 \times 10^{-2}$ & $13.6 \times 10^{-3}$ & - & \citet{Meisenheimer2007}\\
		$2.45 \times 10^{13}$ & $1.45$ & $7.31 \times 10^{-2}$ & $0.4$ & VISIR & \citet{Horst2008}\\
		$2.50 \times 10^{13}$ & $1.52$ & $1.52 \times 10^{-1}$ & $<0.4$ & - & \citet{Asmus2014}\\
		$2.67 \times 10^{13}$ & $9.47 \times 10^{-1}$ & $2.92 \times 10^{-2}$ & $0.4$ & VISIR & \citet{Horst2008}\\
		$2.86 \times 10^{13}$ & $6.42 \times 10^{-1}$ & $2.66 \times 10^{-2}$ & $0.4$ & VISIR & \citet{Horst2008}\\
		$3.61 \times 10^{13}$ & $3.4 \times 10^{-1}$ & $5.0 \times 10^{-2}$ & $13.6 \times 10^{-3}$ & - & \citet{Meisenheimer2007}\\
		$2.42 \times 10^{17}$ & $6.3 \times 10^{-5}$ & $1.8 \times 10^{-5}$ & - & Chandra &\citet{ Evans2006}\\
		$1.33 \times 10^{18}$ & $1.23 \times 10^{-5}$ & $2.25 \times 10^{-7}$ & - & Chandra &\citet{Evans2004}\\
%		$3.02 \times 10^{17}$ & $7.72 \times 10^{-7}$ & - & - & Suzaku & \citet{Markowitz2007}\\
%		$1.33 \times 10^{18}$ & $1.13 \times 10^{-5}$ & - & - & Suzaku & \citet{Markowitz2007}\\
%		$1.45 \times 10^{18}$ & $1.46 \times 10^{-5}$ & - & - & Suzaku & \citet{Markowitz2007}\\
		$6.84 \times 10^{18}$ & $4.50 \times 10^{-6}$ & $8.85 \times 10^{-8}$ & - & BATSE & \citet{Harmon2004}\\
%		$1.06 \times 10^{19}$ & $6.89 \times 10^{-6}$ & - & - & Suzaku & \citet{Markowitz2007}\\
		$1.28 \times 10^{19}$ & $2.09 \times 10^{-6}$ & $3.63 \times 10^{-8}$ & - & BATSE & \citet{Harmon2004}\\
%		$1.45 \times 10^{19}$ & $4.41 \times 10^{-6}$ & - & - & Suzaku & \citet{Markowitz2007}\\
		$2.55 \times 10^{19}$ & $1.72 \times 10^{-6}$ & $2.53 \times 10^{-8}$ & - & BATSE & \citet{Harmon2004}\\
%		$3.57 \times 10^{19}$ & $2.02 \times 10^{-6}$ & - & - & Suzaku & \citet{Markowitz2007}\\
		$6.34 \times 10^{19}$ & $7.20 \times 10^{-7}$ & $3.39 \times 10^{-8}$ & - & BATSE & \citet{Harmon2004}\\
		\hline
	\end{tabular}
\end{table*}

\begin{table*}
	\centering
	\caption{Data table for M84. Columns are same as for Table -[\ref{tab:CenA}]. The table also includes data from the VLBA calibrator database (\url{http://www.vlba.nrao.edu/astro/calib/}) and the ALMA flux calibrator database (\url{https://almascience.eso.org/sc/}).}
	\label{tab:M84}
	\begin{tabular}{lccccr} % four columns, alignment for each
		\hline
		$\nu$ (Hz) & Flux (Jy) & Error (Jy) & Resolution (arcsec) & Instrument & Reference\\
		\hline
		$2.3 \times 10^{9}$ & $1.16 \times 10^{-1}$ & - & $5 \times 10^{-3}$ & VLBA calibrator database & -\\
		$5.0 \times 10^9$ & $1.6 \times 10^{-1}$ & - & $11.0\times 10^{-3}$  & VLBA/VLBI & \citet{Nagar2005}\\							
		$8.4 \times 10^9$ & $1.65 \times 10^{-1}$ & $6.0 \times 10^{-5}$ & $0.5$ & VLA & \citet{Nagar2001}\\
		$8.6 \times 10^{9}$ & $1.36 \times 10^{-1}$ & - & $15\times 10^{-3}$ & VLBA calibrator database & -\\
		$1.5 \times 10^{10}$ & $1.81 \times 10^{-1}$ & - & $2-10\times 10^{-3}$ & VLA & \citet{Nagar2005}\\
		$1.5 \times 10^{10}$ & $1.65 \times 10^{-1}$ & $1.7 \times 10^{-4}$ & $0.15$ & VLA & \citet{Nagar2001}\\
		$1.5 \times 10^{10}$ & $1.67 \times 10^{-1}$ & $1.7 \times 10^{-4}$ & $0.5$ & VLA & \citet{Nagar2001}\\
		$4.3 \times 10^{10}$ & $1.00 \times 10^{-1}$ & $1.0 \times 10^{-2}$ & - & VLBA & \citet{Ly2004}\\
		$9.75 \times 10^{10}$ & $1.16 \times 10^{-1}$ & - & $1$ & ALMA flux calibrator database & -\\
		$2.21 \times 10^{11}$ & $1.19 \times 10^{-1}$ & $1.6 \times 10^{-3}$ & - & SMA (30.1.2016) & \citet{Bower2017}\\
		$2.21 \times 10^{11}$ & $1.29 \times 10^{-1}$ & $1.9 \times 10^{-3}$ & - & SMA (21.2.2016) & \citet{Bower2017}\\
		$2.21 \times 10^{11}$ & $1.60 \times 10^{-1}$ & $1.5 \times 10^{-3}$ & - & SMA (29.1.2016) & \citet{Bower2017}\\
		$2.31 \times 10^{11}$ & $1.14 \times 10^{-1}$ & - & $1$ & ALMA flux calibrator database & -\\
		$2.33 \times 10^{11}$ & $1.14 \times 10^{-1}$ & $1.4 \times 10^{-3}$ & - & SMA (30.1.2016) & \citet{Bower2017}\\
		$2.33 \times 10^{11}$ & $1.45 \times 10^{-1}$ & $1.8 \times 10^{-3}$ & - & SMA (21.2.2016) & \citet{Bower2017}\\
		$2.33 \times 10^{11}$ & $1.71 \times 10^{-1}$ & $1.6 \times 10^{-3}$ & - & SMA (29.2.2016) & \citet{Bower2017}\\
		$2.7 \times 10^{13}$ & $<8.0 \times 10^{-3}$ & - & - & - & \citet{Asmus2014}\\
		$1.87 \times 10^{14}$ & $8.51 \times 10^{-5}$ & $6.0 \times 10^{-5}$ & $0.5$ & HST & \citet{Buttiglione2009}\\
		$3.02 \times 10^{17}$ & $1.76 \times 10^{-8}$ & - & - & Chandra & \citet{Gonzales2009}\\
		$6.65 \times 10^{17}$ & $2.26 \times 10^{-8}$ & $1.2 \times 10^{-8}$ & - & Chandra & \citet{Balmaverde2006}\\
		$1.00 \times 10^{18}$ & $1.40 \times 10^{-8}$ & - & - & Chandra & \citet{Donato2004}\\
		$1.45 \times 10^{18}$ & $5.59 \times 10^{-9}$ & - & - & Chandra & \citet{Gonzales2009}\\
		\hline
	\end{tabular}
\end{table*}

\begin{table*}
	\centering
	\caption{Data table for NGC 4594. Columns are same as for Table -[\ref{tab:CenA}]}
	\label{tab:NGC4594}
	\begin{tabular}{lccccr} % four columns, alignment for each
		\hline
		$\nu$ (Hz) & Flux (Jy) & Error (Jy) & Resolution (arcsec) & Instrument & Reference\\
		\hline
		$1.4 \times 10^9$ & $5.96 \times 10^{-2}$ & $6.0 \times 10^{-3}$ & $(23 \times 6) \times 10^{-3}$  & VLBA & \citet{Hada2013}\\
		$2.3 \times 10^9$ & $6.21 \times 10^{-2}$ & $6.2 \times 10^{-3}$ & $(9 \times 4)\times 10^{-3}$ & VLBA & \citet{Hada2013}\\
		$5.0 \times 10^9$ & $7.43 \times 10^{-2}$ & $7.4 \times 10^{-3}$ & $(4 \times 1)\times 10^{-3}$ & VLBA & \citet{Hada2013}\\
		$8.4 \times 10^9$ & $8.02 \times 10^{-2}$ & $8.0 \times 10^{-3}$ & $(2 \times 1)\times 10^{-3}$ & VLBA & \citet{Hada2013}\\
		$1.52 \times 10^{10}$ & $8.71 \times 10^{-2}$ & $8.7 \times 10^{-3}$ & $(1 \times 0.5)\times 10^{-3}$ & VLBA & \citet{Hada2013}\\
		$2.38 \times 10^{10}$ & $8.80 \times 10^{-2}$ & $8.8 \times 10^{-3}$ & $(1 \times 0.4)\times 10^{-3}$ & VLBA & \citet{Hada2013}\\
		$4.32 \times 10^{10}$ & $9.10 \times 10^{-2}$ & $9.1 \times 10^{-3}$ & $(0.8 \times 0.2)\times 10^{-3}$ & VLBA & \citet{Hada2013}\\
		$2.5\times 10^{11}$ & $1.85 \times 10^{-1}$ & $2.0 \times 10^{-3}$ & - & MAMBO & \citet{Vlahakis2008}\\
		$3.45 \times 10^{11}$ & $2.3 \times 10^{-1}$ & $3.5 \times 10^{-2}$  & - & LABOCA & \citet{Krause2006}\\
		$3.45 \times 10^{11}$ & $2.42 \times 10^{-1}$ & $3.0 \times 10^{-3}$ & - & LABOCA & \citet{Vlahakis2008}\\
		$3.53 \times 10^{11}$ & $2.5 \times 10^{-1}$ & $6.0 \times 10^{-2}$ & - & SCUBA & \citet{Krause2006}\\
		$1.25 \times 10^{13}$ & $1.04 \times 10^{-2}$ & - & - & - & \citet{YongShi2010}\\
		$2.4 \times 10^{13}$ & $4.0 \times 10^{-3}$ & $1.0 \times 10^{-3}$ & - & GeminiS & \citet{Asmus2014}\\
		$2.5 \times 10^{13}$ & $1.3 \times 10^{-3}$ & - & - & - & \citet{YongShi2010}\\
		$2.54 \times 10^{13}$ & $< 7.0 \times 10^{-3}$ & - & - & VLT & \citet{Reunanen2010}\\
		$9.09\times 10^{14}$ & $5.55 \times 10^{-5}$ & - & - & HST & \citet{Maoz2005}\\
		$1.2 \times 10^{15}$ & $1.57 \times 10^{-5}$ & - & - & HST & \citet{Maoz2005}\\
		$3.02 \times 10^{17}$ & $1.82 \times 10^{-7}$ & - & - & - & \citet{Pellegrini2003}\\
		$3.02 \times 10^{17}$ & $7.45 \times 10^{-8}$ & - & - & Chandra & \citet{Gonzales2009}\\
		$1.0 \times 10^{18}$ & $6.99 \times 10^{-8}$ & $9.8 \times 10^{-10}$ & - & Chandra & \citet{Grier2011}\\
		$1.45 \times 10^{18}$ & $5.53 \times 10^{-8}$ & - & - & Chandra & \citet{Gonzales2009}\\
		\hline
	\end{tabular}
\end{table*}

\begin{table*}
	\centering
	\caption{Data table for NGC 3998. Columns are same as for Table -[\ref{tab:CenA}]}
	\label{tab:NGC3998}
	\begin{tabular}{lccccr} % four columns, alignment for each
		\hline
		$\nu$ (Hz) & Flux (Jy) & Error (Jy) & Resolution (arcsec) & Instrument & Reference\\
		\hline
		$1.50 \times 10^{10}$ & $5.7 \times 10^{-2}$ & -  & - & VLA & \citet{Nagar2005}\\
		$1.67 \times 10^{13}$ & $1.5 \times 10^{-1}$ & $1.52 \times 10^{-2}$ & $0.4$ & GeminiN & \citet{Asmus2014}\\
		$2.5\times 10^{13}$ & $7.1 \times 10^{-2}$ & $8.0 \times 10^{-3}$ & $0.4$ & GeminiN & \citet{Asmus2014}\\
		$3.8 \times 10^{14}$ & $1.1 \times 10^{-3}$ & - & $0.2$ & HST & \citet{Delgado2008}\\
		$3.8 \times 10^{14}$ & $1.4 \times 10^{-3}$ & - & $0.1$ & XMM-Newton & \citet{Ptak2004}\\
		$5.47 \times 10^{14}$ & $1.6 \times 10^{-3}$ & - & $0.2$ & HST & \citet{Delgado2008}\\
		$5.5 \times 10^{14}$ & $8.0 \times 10^{-4}$ & - & $0.1$ & XMM-Newton & \citet{Ptak2004}\\
		$8.7 \times 10^{14}$ & $1.2 \times 10^{-3}$ & - & $3$ & XMM-Newton & \citet{Ptak2004}\\
		$9.09\times 10^{14}$ & $5.54 \times 10^{-4}$ & - & - & HST & \citet{Maoz2005}\\
		$1.2 \times 10^{15}$ & $4.14 \times 10^{-4}$ & - & - & HST & \citet{Maoz2005}\\
		$1.4\times 10^{15}$ & $5.6 \times 10^{-4}$ & - & $2$ & XMM-Newton & \citet{Ptak2004}\\
		$3.02 \times 10^{17}$ & $1.04 \times 10^{-6}$ & - & - & Chandra & \citet{Gonzales2009}\\
		$1.45 \times 10^{18}$ & $5.69 \times 10^{-7}$ & - & - & Chandra & \citet{Gonzales2009}\\
		\hline
	\end{tabular}
\end{table*}

\begin{table*}
	\centering
	\caption{Data table for NGC 4278. Columns are same as for Table -[\ref{tab:CenA}]}
	\label{tab:NGC4278}
	\begin{tabular}{lccccr} % four columns, alignment for each
		\hline
		$\nu$ (Hz) & Flux (Jy) & Error (Jy) & Resolution (arcsec) & Instrument & Reference\\
		\hline
		$5.0 \times 10^9$ & $1.35 \times 10^{-1}$ & - & $2.5\times 10^{-3}$ & VLBA/VLA & \citet{Giroletti2005}\\
		$5.0 \times 10^9$ & $3.73 \times 10^{-2}$ & -  & $10 \times 10^{-3}$ & VLBA/VLBI & \citet{Nagar2005}\\
		$8.4 \times 10^9$ & $9.49 \times 10^{-2}$ & - & $(3.2 \times 1.9) \times10^{-3}$ & VLA & \citet{Giroletti2005}\\
		$1.5 \times 10^{10}$ & $7.99 \times 10^{-2}$ & - & $0.15$ & VLA & \citet{Nagar2001}\\
		$1.5 \times 10^{10}$ & $8.83 \times 10^{-2}$ & - & $(2-10)\times 10^{-3}$ & VLA & \citet{Nagar2005}\\
		$2.2 \times 10^{10}$ & $7.3 \times 10^{-2}$ & - & $<1$ & VLA & \citet{Giroletti2005}\\
		$2.68 \times 10^{13}$ & $2.9 \times 10^{-3}$ & $1.0 \times 10^{-3}$ & - & VLT & \citet{Asmus2014}\\
		$3.32 \times 10^{14}$ & $2.36 \times 10^{-4}$ & - & $0.15$ & HST (Dec 2006) & \citet{Younes2010}\\
		$3.32 \times 10^{14}$ & $2.55 \times 10^{-4}$ & - & $0.15$ & HST (Jan 2007) & \citet{Younes2010}\\
		$6.32 \times 10^{14}$ & $1.19 \times 10^{-4}$ & - & $0.15$ & HST (Dec 2006) & \citet{Younes2010}\\
		$6.32 \times 10^{14}$ & $1.44 \times 10^{-4}$ & - & $0.15$ & HST (Jan 2007) & \citet{Younes2010}\\
		$1.30 \times 10^{15}$ & $2.79 \times 10^{-5}$ & - & $3.0$ & XMM-Newton (May 2004) & \citet{Younes2010}\\
		$1.21 \times 10^{17}$ & $3.27 \times 10^{-7}$ & - & $3.34$ & Chandra (Feb 2007) & \citet{Younes2010}\\
		$3.02 \times 10^{17}$ & $6.95 \times 10^{-8}$ & - & - & Chandra & \citet{Gonzales2009}\\
		$1.45 \times 10^{18}$ & $1.18 \times 10^{-8}$ & - & - & Chandra & \citet{Gonzales2009}\\
		$1.93 \times 10^{18}$ & $1.22 \times 10^{-8}$ & - & $3.34$ & Chandra (Feb 2007) & \citet{Younes2010}\\
		\hline
	\end{tabular}
\end{table*}


%%%%%%%%%%%%%%%%%%%%%%%%%%%%%%%%%%%%%%%%%%%%%%%%%%

% Don't change these lines
\bsp	% typesetting comment
\label{lastpage}
\end{document}